\documentclass[english]{sig-alternate-10pt}
\usepackage[T1]{fontenc}
\usepackage[latin9]{inputenc}
\usepackage{array}
\usepackage{float}
\usepackage{multirow}
\usepackage{graphicx, tabularx}
\usepackage{url}

\usepackage[
  pass,
]{geometry}

\makeatletter

\providecommand{\tabularnewline}{\\}

\@ifundefined{showcaptionsetup}{}{%
 \PassOptionsToPackage{caption=false}{subfig}}
\usepackage{subfig}
\makeatother

\begin{document}

\title{Measurement-Based Modelling of LTE Performance in Dublin City}

\author{Miguel B\'{a}guena Albaladejo$^1$, Douglas J.Leith$^2$\thanks{DL was supported by SFI grants 11/II/1177 and 13/RC/2077.}, Pietro Manzoni$^1$\\$^1$Universitat Polit\`{e}cnica de Val\'{e}ncia, Spain $^2$Trinity College Dublin, Ireland}
\maketitle
\begin{abstract}
LTE/4G is the next generation of cellular network which specifically aims to improve the network performance for data traffic and is currently being rolled out by many network operators.   We present results from an extensive LTE measurement campaign in Dublin, Ireland using a custom performance measurement tool.   Performance data was measured at a variety of locations within the city (including cell edge locations, indoors, outdoors etc) as well as for mobile users on public transport within the city.   Using this data we derive a model of the characteristics of link layer RTT and bandwidth vs link signal strength.   This model is suited to use for performance evaluation of applications and services, and since it is based on real measurements it allows realistic evaluation of performance.
\end{abstract}

\section{Introduction}

LTE/4G has undergone rapid rollout and now plays a major role in cellular communications, particularly within urban areas.  The LTE link layer is, however, complex and contains numerous design parameters the choice of which is left proprietary.  The complexity of the LTE MAC makes simulation computationally demanding, while the many unspecified design parameters make selection of realistic configurations problematic.    Further, although there have been a number of measurement studies of transport and application layer performance over LTE, measurement studies of actual LTE link layer behaviour in the field are much more limited.

Motivated by these observations, in this paper we present results from a large-scale LTE measurement campaign carried out in Dublin, Ireland.    These measurements provide an accurate snapshot of the service actually offered to the users and the measurement tools developed therefore potentially provide a valuable source of information for operators and users.

Using this measurement data we derive an empirical model of the characteristics of link layer RTT and bandwidth, taking account of the changes in the distribution of RTT and bandwidth with link quality and choice of network operator.    
We find that the RTT over an LTE hop is distributed quite differently from other link layers (such as WiFi) and can be modelled as a mixture of Gaussians.  This model captures the scheduling granularity for HARQ \emph{etc} within the LTE link layer in a new and useful way.   The model is predictive: for a given link quality a model RTT distribution can be calculated and used to generate realistic link layer RTT time series.   We find that the bandwidth over an LTE hop increases in variability as the link quality, and so transmit rate, improves.  It seems likely that this is associated with the time granularity of LTE link layer scheduling.   We present a measurement-based model of the dependence of bandwidth on link quality that, again, is predictive in nature and allows generation of realistic bandwidth time series.

Our hope is that this model will assist the research community in realistic, yet reproducible, performance evaluation of new tools and applications.  Since this empirical model is much simpler than detailed LTE simulation models, it should be more convenient for evaluation of application layer performance and for evaluation over longer time scales (where simulation would take too long).   Since the model is based on measurements from actual LTE network deployments, it reflects realistic configurations.


\section{Related work} \label{sec:related-work}

Since LTE is a relatively new technology, the existing literature on performance of cellular networks mostly deals with earlier or alternative cellular technologies \cite{Tso:2010:MDS:1860093.1860105:p81-tso,Baltrunas:2014:MRM:2663716.2663725:IMC2014-p45,Gember:2012:OIM:2398776.2398807:p287-gember} and on particular measurements of specific physical-level characteristics or events \cite{Merz:2014:PLH:2627585.2627589:p47-merz,Park:2014:PAL:2584567.2584584:779_2013_Article_674,Nguyen:2014:USM:2736222.2736236:p1559-nguyen,Huang:2012:CEP:2307636.2307658:1_huang_Mobisys2012}.

A number of other studies measure throughput and/or latency information, but tend to focus on their impact on TCP or its derivatives \cite{Huang:2013:ISL:2486001.2486006:sigcomm2013-p363,Deng:2014:WLB:2663716.2663727:IMC2014-p181,Shafiq:2013:FLC:2494232.2465754:p17-shafiq,Xu:2012:IIT:2387238.2387248:p39-xu,Nguyen:2014:TUT:2627585.2627594:p41-nguyen,Chen:2013:MSM:2504730.2504751:p455-chen,Shafiq:2014:UIN:2591971.2591975:p367-shafiq}.  Additionally, there are studies that present measurements but do not use these to develop a model suited to performance evaluation \cite{Sommers:2012:CVW:2398776.2398808:p301-sommers}.

There has been work on simulation \cite{Piro:2011:LMN:2151054.2151129:p415-piro,lena,simulte,4GSim} and modelling \cite{Gupta:2014:DLB:2643230.2643247:p91-Gupta,Hsu:2010:MVD:1837274.1837282:p18-hsu,Jovanovic:2014:QNP:2653481.2655236:p25-jovanovic,Shafiq:2011:CMI:1993744.1993776:p305-shafiq}.  However, such tools may be of limited utility because they do not provide the emulation needed to evaluate real-time aspects of actual applications, the models may be overly specific, or they may be at too low layer within the network stack.

Our study is different from the aforementioned work in a number of way, including because it focuses on LTE, it studies performance independently from TCP, and because our explicit aim is to build realistic yet simple models useful for evaluating a wide range of protocol and application performance. Since information is presented in terms of commonly used performance indicators (bandwidth and delay), channels with the approriate characteristics can be readlity emulated for testing of real applications now available on the market.

\section{Experimental setup} \label{sec:setup}

\textbf{Hardware}. We use two Asus EeePC 4G (Intel Celeron M 353 -
900MHz, 512 MB RAM) laptops as cellular UEs for carrying out measurements, each equipped with a 4G USB modem (Alcatel One Touch L100V).   These run Lubuntu 14.04 and use ModemManager
v1.0.0 to connect to the cellular network. This tool uses libqmi (1.4.0) to operate the modem and it was also used to set the operating mode of the modem to 4G. The qmi\_wwan driver in the Linux kernel 3.16.0\_30 manages the modem. The modem internal state is gathered by using the Hayes command set.  GPS is used to measure the UE location. 

\textbf{Measurement Tool}. To measure link characteristics we used a custom tool.  This connects with a remote server located in Trinity College Dublin (which is located in the city centre) and this server tests the downlink connection to the UE every second as follows: 1) in the first 500ms the server sends a burst of 50 UDP packets of 1470 Bytes each back-to-back to the UE 
, which records the packet arrival times and uses these to estimate the downlink bandwidth, and 2) in the second 500ms the server sends a UDP packet every 50ms (10 packets) and the UE's responses are used to estimate the round-trip time (RTT).  Packet losses in this second half are used as source of packet loss information.   All traffic (timestamps, packet size, packet ids) is logged for offline analysis, with the UE and remote server clocks synchronized using NTP.  This tool is intended to be light and simple so as to avoid the active measurements themselves significantly affecting the path characteristics.   Using traceroute we confirmed that the cellular network connects directly with the Irish academic network, with the college having a 10Gbps connection and the server 1Gbps so it is fairly safe to assume that the wireless hop acts as the path bottleneck.  The RTT over the academic part of the path is reported by traceroute to be less than 1ms, so measured RTTs are dominated by delay on the cellular network (which includes the wireless hop plus transit of the operators wired network).  As part of validating the tool we measured path characteristics in a lab testbed using dummynet to modify the link properties in a controlled manner.  The accuracy of the measurement tool is summarised in Table \ref{tab:valid}. 

\begin{figure}
\centering
\includegraphics[width=0.7\columnwidth]{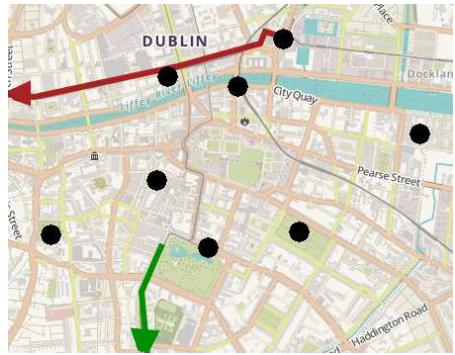}
\caption{Map indicating measurement locations (black dots) and tram lines (in red and green) in Dublin.}\label{fig:map}
\end{figure}

\begin{figure}
\vspace{-0.3cm}
\subfloat[Operator T]{
\includegraphics[width=0.45\columnwidth]{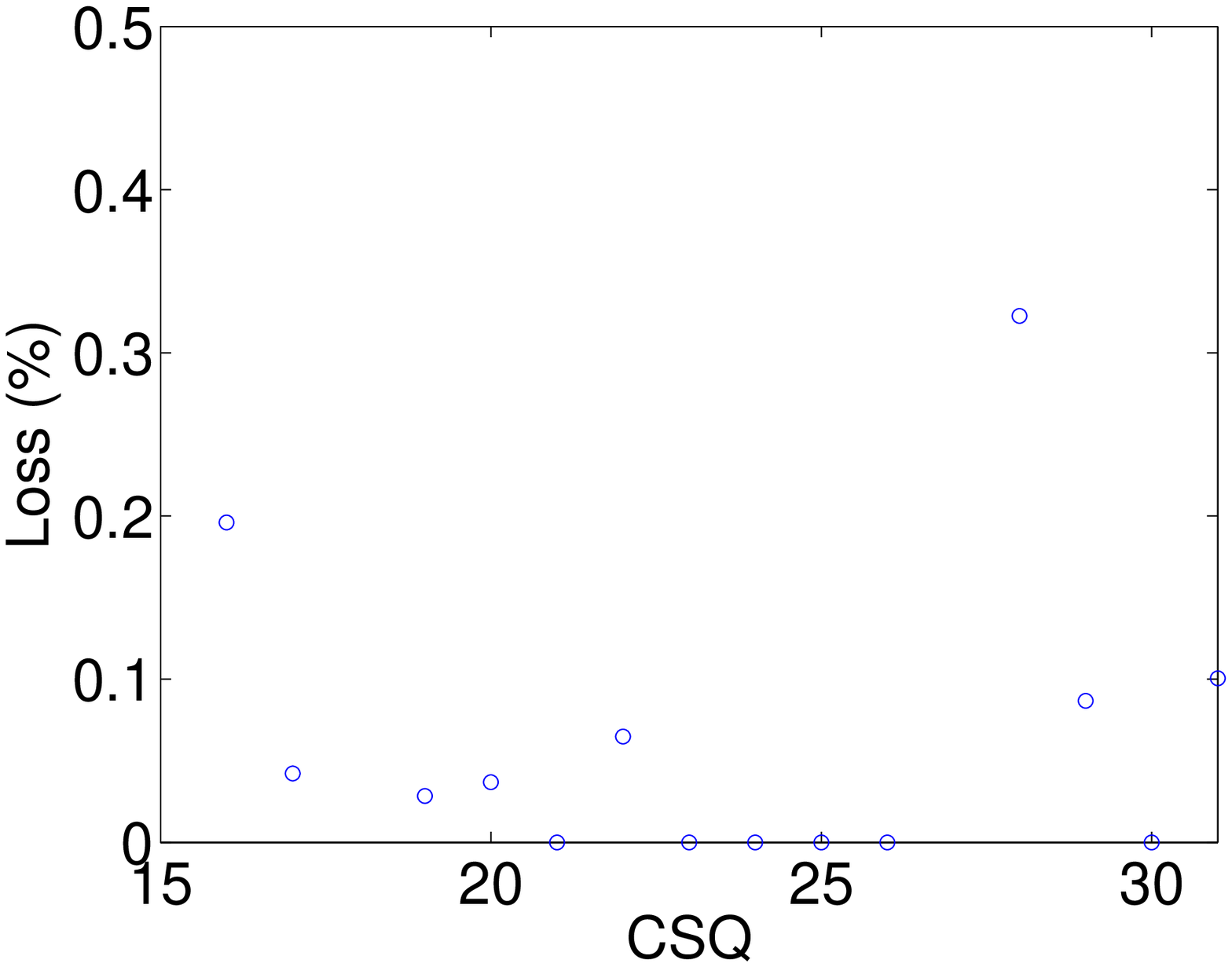}
}
\subfloat[Operator M]{
\includegraphics[width=0.45\columnwidth]{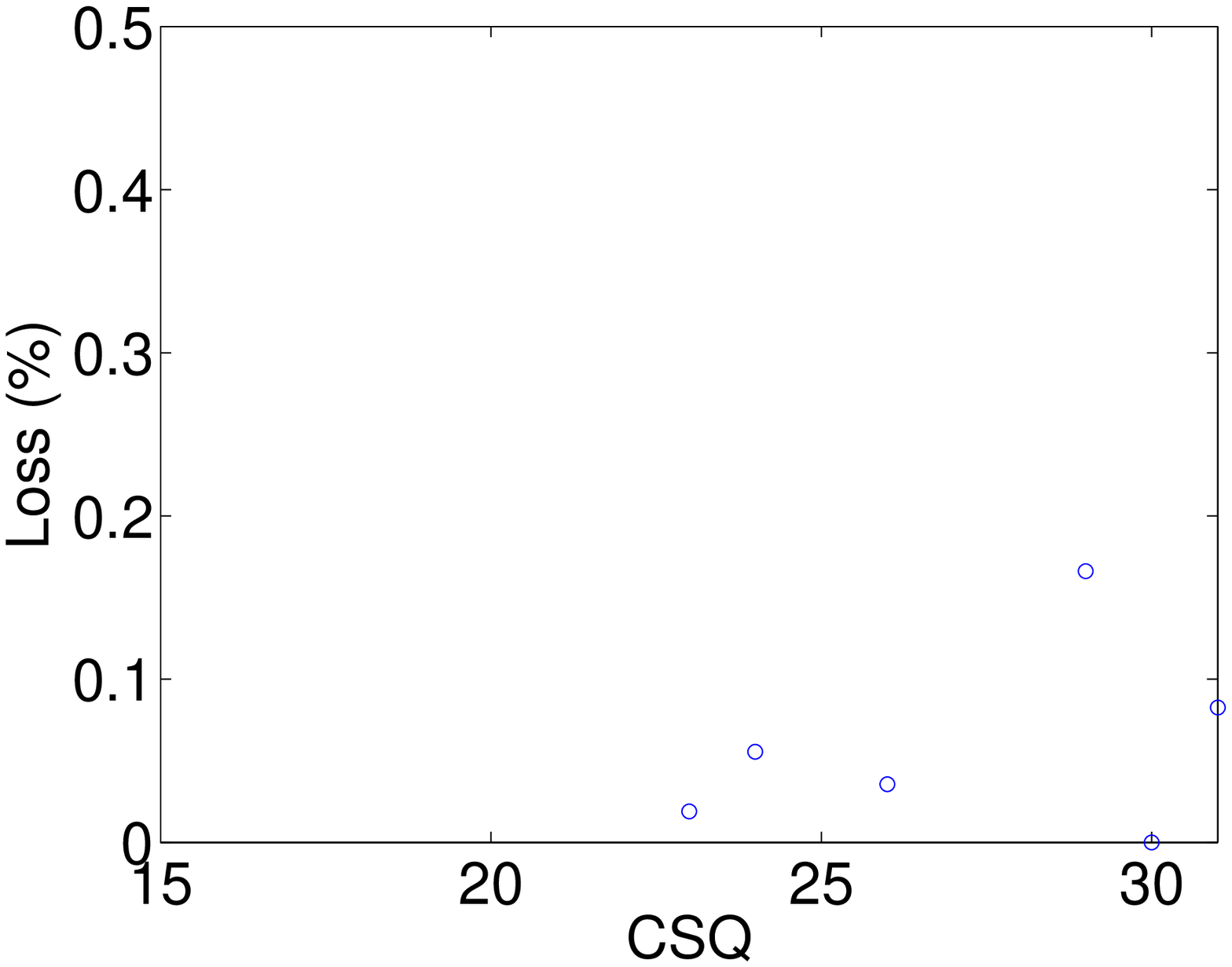}
}
\caption{Measured loss rate vs CSQ. Stationary UE.}\label{fig:loss}
\end{figure}

\textbf{Measurement Campaign}.   Performance data was measured at a variety of locations within the city (including cell edge locations, indoors, outdoors etc) as well as for mobile users on public transport within the city, see Figure \ref{fig:map}. 
These locations covered a large part of the Dublin city centre area, as well as being quite diverse in nature (including parks, public areas and railway stations). The tests were performed over several weeks, during the months of March and April of 2015, and over a range of times.  Measurements were taken for two major mobile operators, which we refer to as T and M here.   We observed low packet loss rates in our tests of less than 0.35\%, see Figure \ref{fig:loss}, and so do not report further on these here. 


\begin{table}
\scriptsize
\centering
\subfloat[][Bandwidth (Mbps)]{
\begin{tabularx}{0.45\columnwidth}{|X|c|c|}
\hline
BW & $\mu$ & $\sigma$  \tabularnewline
\hline 
100 &  99.22 & 7.11 \tabularnewline
\hline 
75 &  74.13 & 4.12  \tabularnewline
\hline 
50 &  48.76 & 1.61  \tabularnewline
\hline 
25 &  24.17 & 0.37  \tabularnewline
\hline 
10 &  9.77 & $\sim0$  \tabularnewline
\hline 
1 &  0.98 & $\sim0$  \tabularnewline
\hline 
0.5 &  0.49 & $\sim0$  \tabularnewline
\hline 
\end{tabularx}
}\quad
\subfloat[][RTT (ms)]{
\begin{tabularx}{0.45\columnwidth}{|X|c|c|}
\hline 
RTT & $\mu$ & $\sigma$  \tabularnewline
\hline 
1000 &  1000.54 & 0.05 \tabularnewline
\hline 
750 & 750.57 & 0.04 \tabularnewline
\hline 
500 &  500.62 & 0.03 \tabularnewline
\hline 
250 &  250.65 & 0.03 \tabularnewline
\hline 
100 &  100.68 & 0.09 \tabularnewline
\hline 
75 &  75.67 & 0.05 \tabularnewline
\hline 
50 &  50.68 & 0.08 \tabularnewline
\hline 
25 &  25.67 & 0.05 \tabularnewline
\hline 
10 &  10.68 & 0.05 \tabularnewline
\hline 
5 &  5.68 & 0.04 \tabularnewline
\hline 
\end{tabularx}
}
\caption{Accuracy of measurement tool, lab measurements ($\mu$ denotes mean value, $\sigma$ standard deviation).}\label{tab:valid}
\normalsize
\end{table}

\section{Delay} \label{sec:delay}

In this section we show that the RTT over the LTE wireless hop can be modelled as i.i.d and following a mixture of Gaussians whose parameters depend on the link quality.   Key to this is the observation that UE traffic is queued separately at the LTE basestation and so link layer RTT can be inferred despite fluctuations in cell traffic load.

\subsection{Per-UE Queueing at Basestation}

\begin{figure}
\subfloat[][UE downloading file]{\includegraphics[width=0.465\columnwidth]{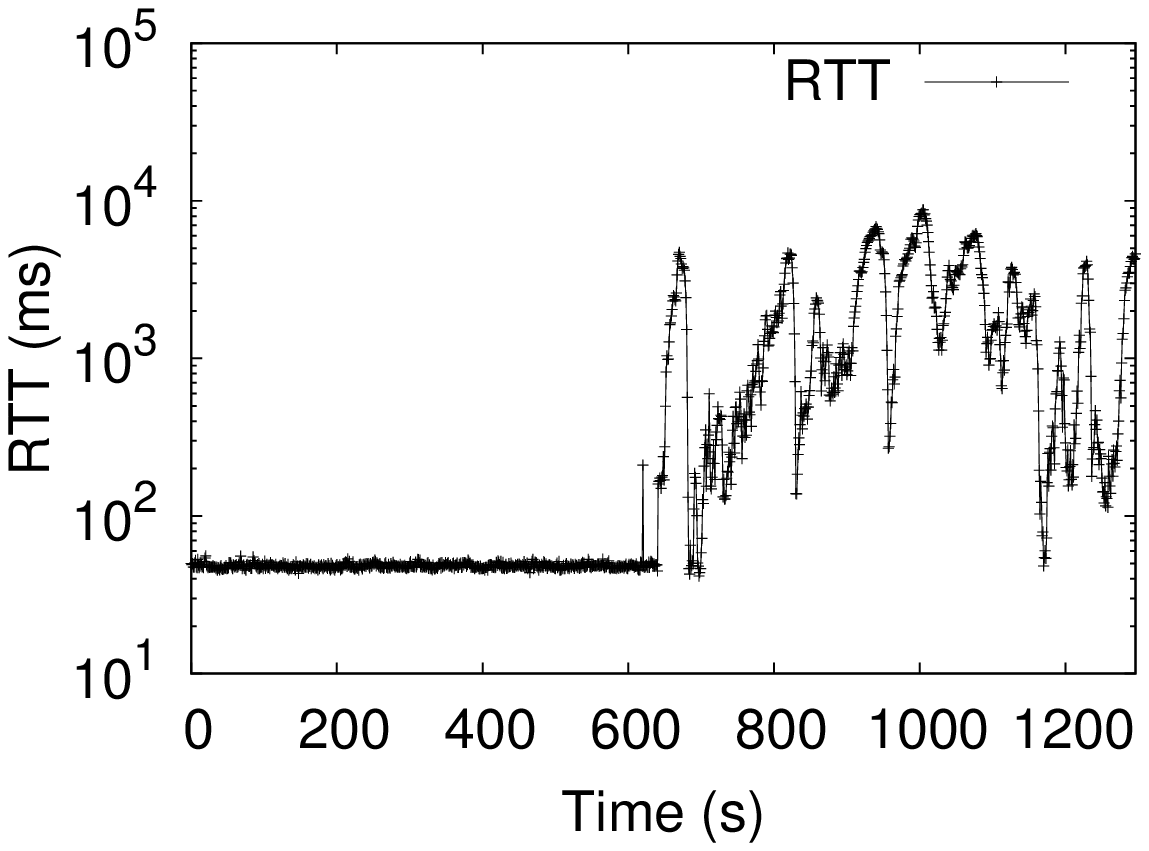}
}
\quad
\subfloat[][Other UE]{\includegraphics[width=0.465\columnwidth]{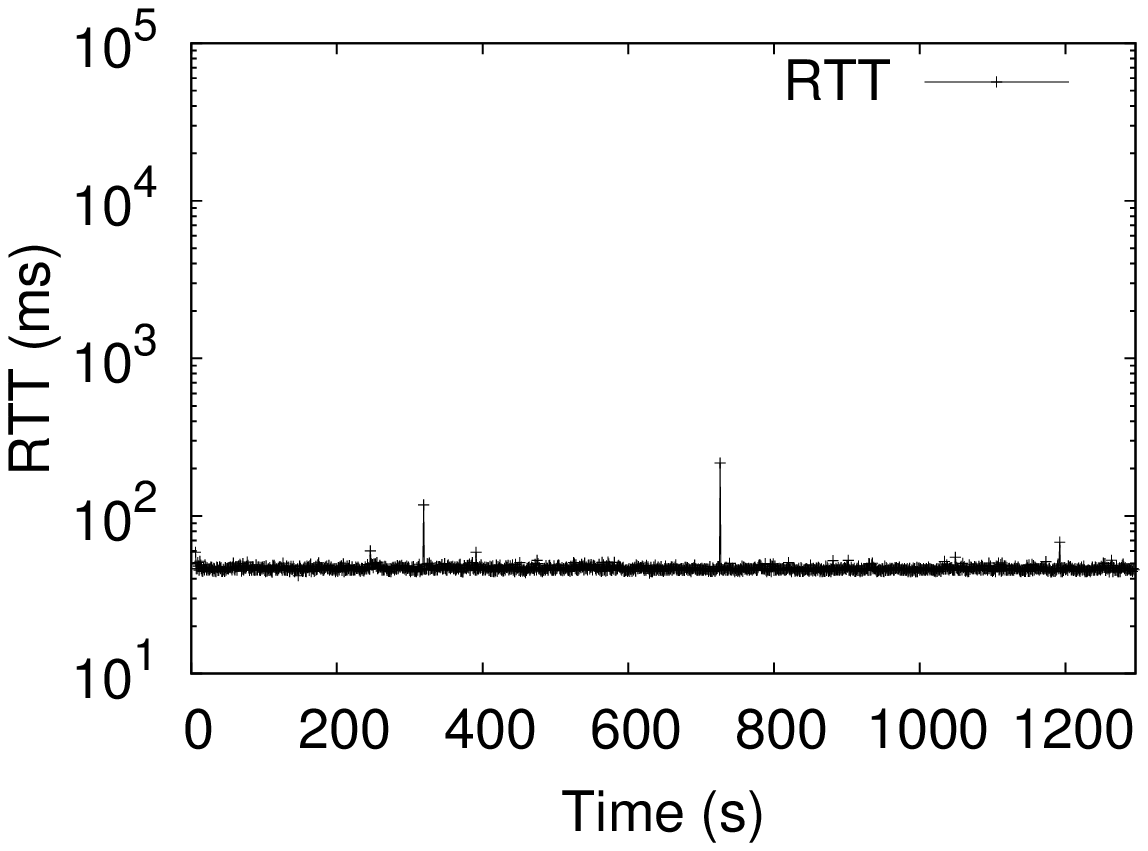}
}\caption{RTT measured at two UEs while one is downloading a large file.  Both UEs connected to the same network cell and located beside one another.}\label{fig:Separability-of-delay}
\end{figure}

We placed two UEs side by side.  On one UE we then started a large file download while simultaneously measuring the bandwidth and RTT on both UEs.   Figure \ref{fig:Separability-of-delay} plots the RTTs measured by the two UEs.   It can be seen that on the UE carrying out the download the measured RTT increases from around 40ms to greater than 1000ms after the download starts.   In contrast, for the second UE the RTT is unchanged.  The data shown is representative of that measured at a number of locations.  We infer that traffic for different UEs is queued is queued separately at the cellular basestation, with the inceased delay measured by the downloading UE being due to queueing delay at the basestation downlink for that UE.   This also indicates that RTT measurements made by an unloaded UE (\emph{i.e.} one which is not generating enough traffic to cause queueing) can be expected to be insensitive to network load and so potentially provide insight into the characteristics (scheduler latency, ARQ \emph{etc}) of the wireless link layer.

%

\subsection{RTT vs Link Quality: Operator T}

\begin{figure}
\subfloat[Mean]{
\includegraphics[width=0.45\columnwidth]{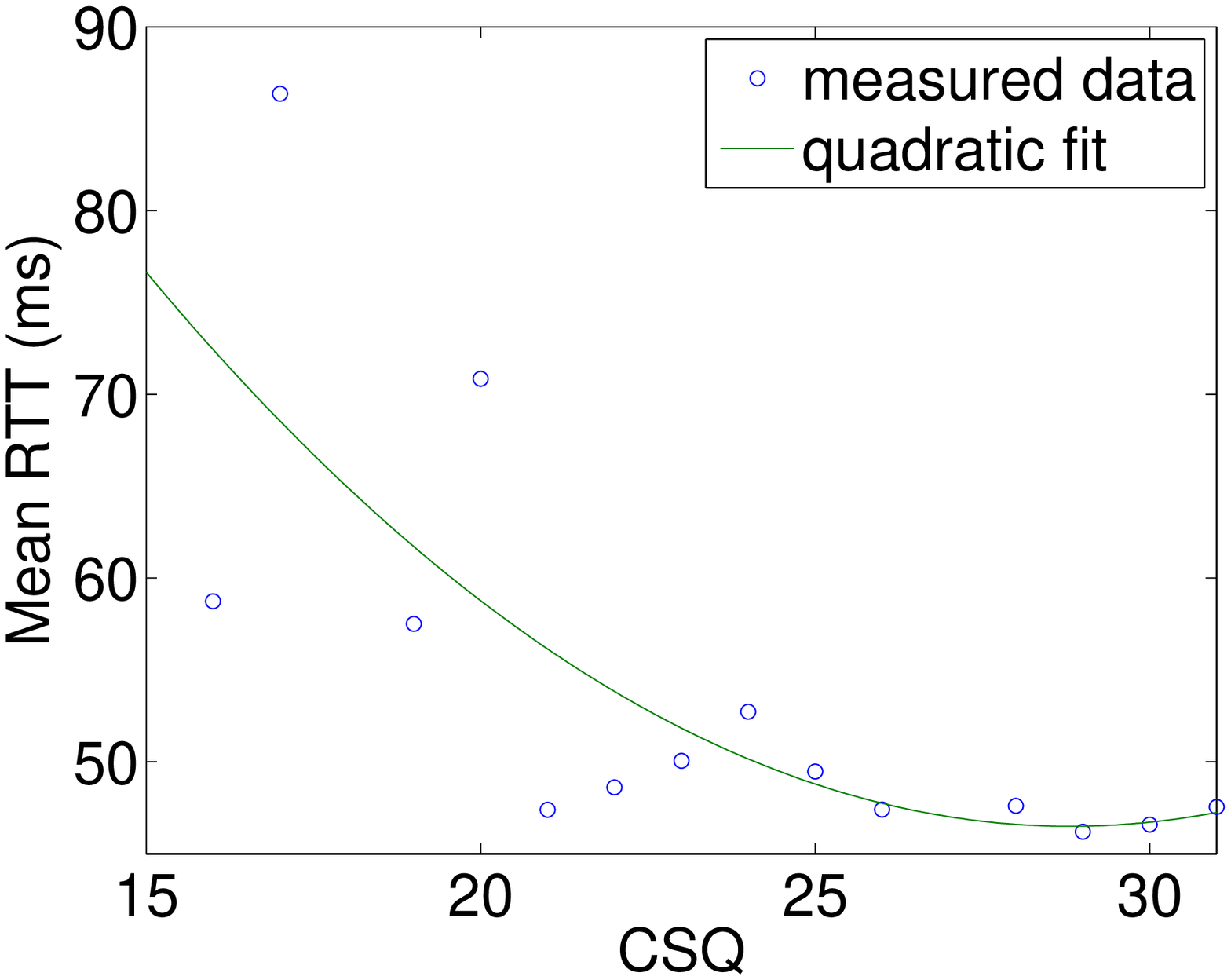}
}\quad
\subfloat[Standard deviation]{
\includegraphics[width=0.45\columnwidth]{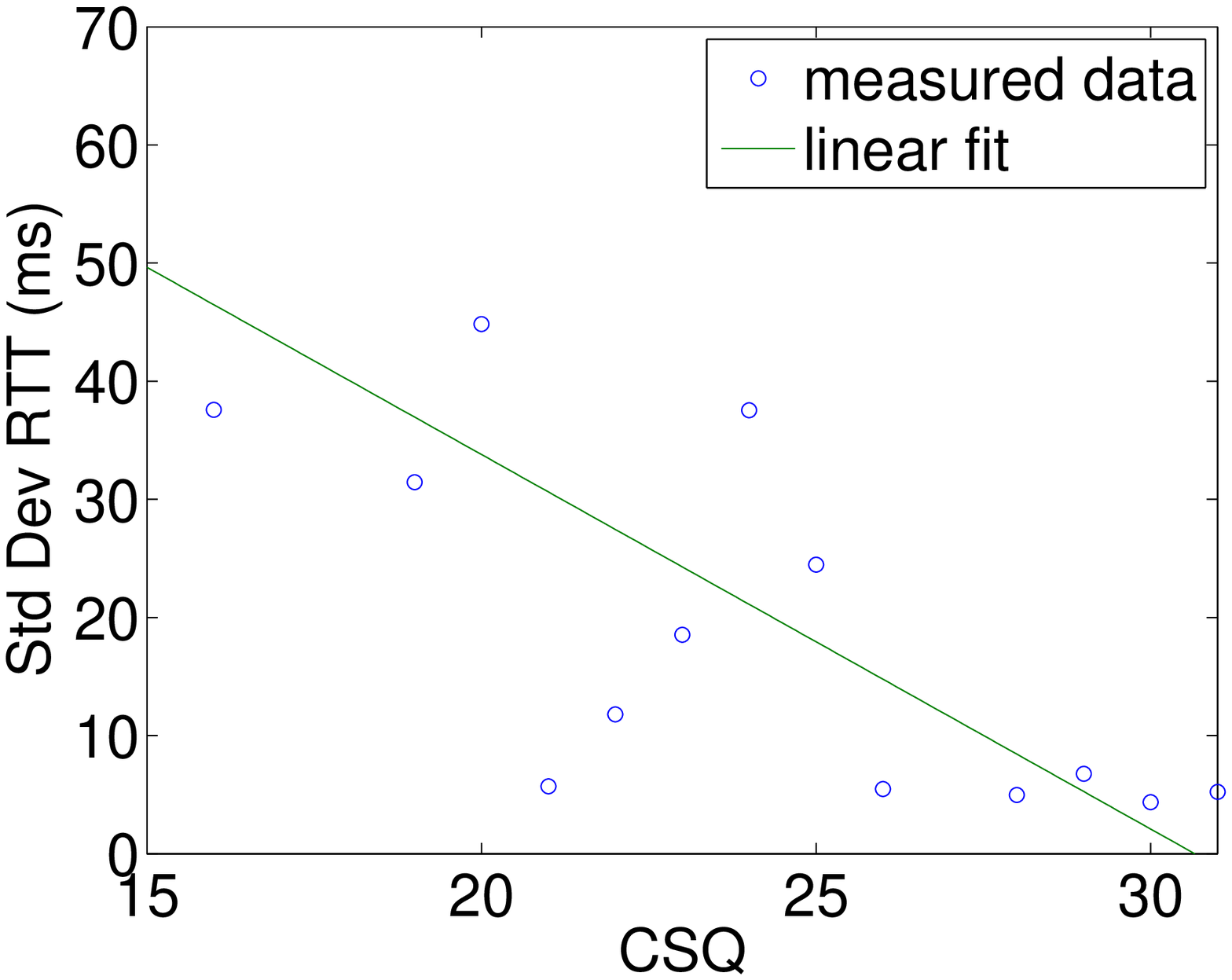}
}
\caption{\label{fig:RTT-mean-distribution}RTT mean and standard deviation, plus quadratic and linear fits respectively. Stationary UE, operator T.}
\end{figure}

The UE modem reports a CSQ value which indicates the RSSI on the wireless downlink. According to \cite{ATCommandSet}, this scale goes from 0, which is equivalent to -113 dbm (or less), and ends on 31, which is equivalent to -51 (or greater).    In our measurements we observed CSQ values in the range 10-31.   Figure \ref{fig:RTT-mean-distribution} plots the mean and standard deviation of RTT vs CSQ for a stationary UE.   This data was collected from six different locations in order to obtain a range of link qualities.   It can be seen that the mean delay increases slowy as the CSQ (and so RSSI) falls, and similarly for the standard deviation of delay.  The mean and standard deviation variation with CSQ can be modeled by, respectively, $\mu=177.69-9.11\times CSQ+0.158\times CSQ^2$ and $\sigma=97.21-3.17\times CSQ$ (these fits are indicated on Figure \ref{fig:RTT-mean-distribution}). 

\begin{figure}
\centering
\subfloat[Distribution, CSQ 19]{
\includegraphics[width=0.45\columnwidth]{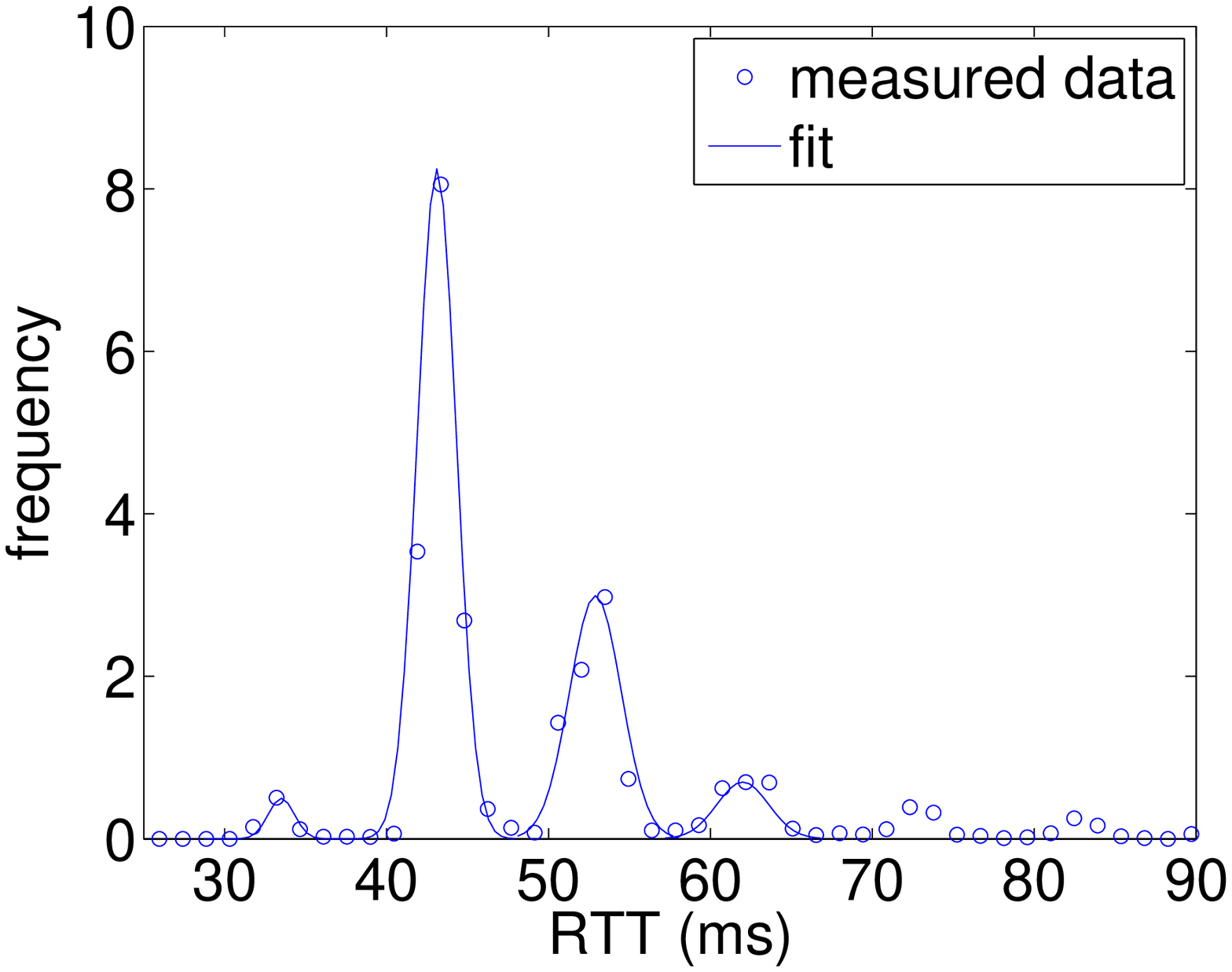}
}\quad
\subfloat[Distribution, CSQ 31]{
\includegraphics[width=0.45\columnwidth]{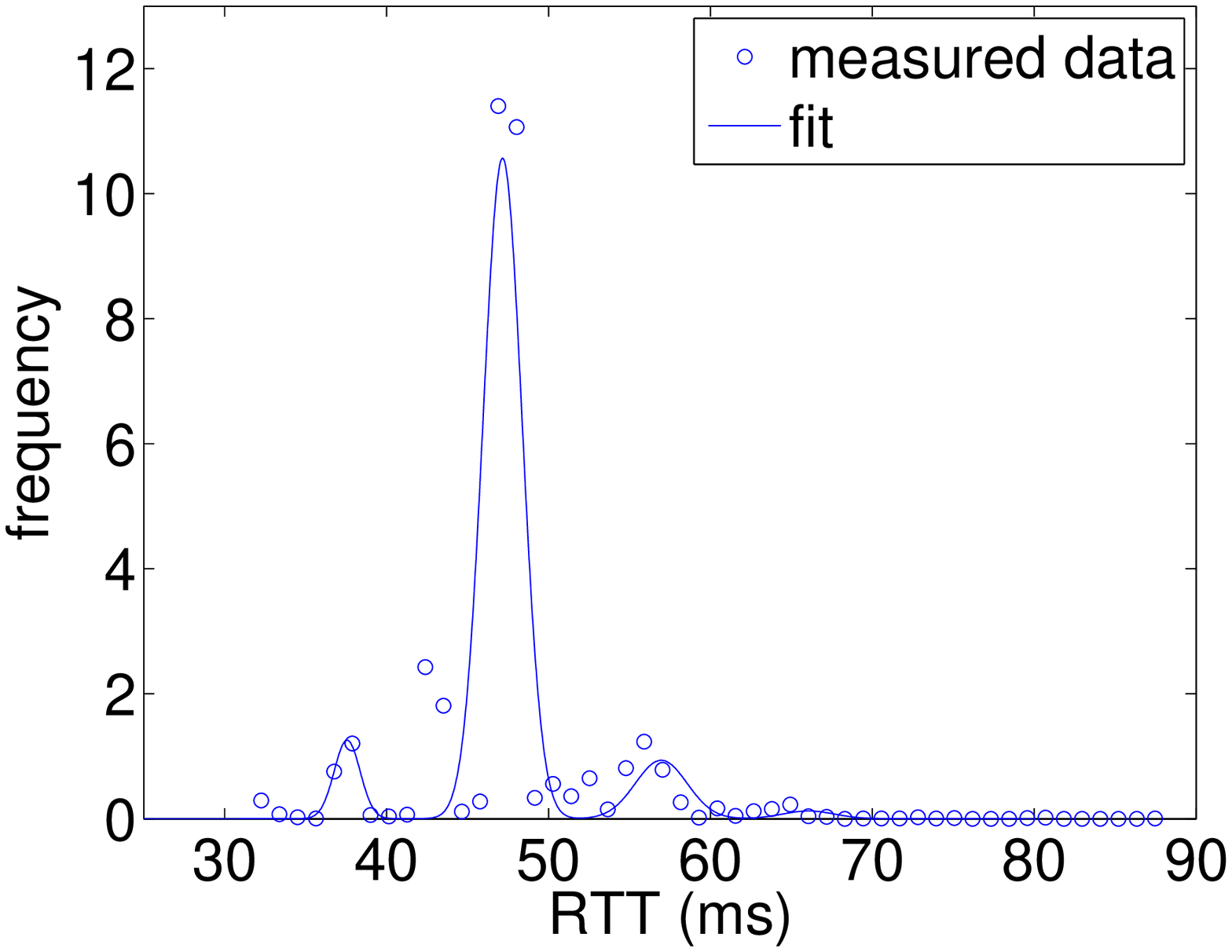}
}
\caption{\label{fig:Link-layer-RTT-fit}Measured RTT.  Stationary UE, operator T. }
\end{figure}

In more detail, Figure \ref{fig:Link-layer-RTT-fit} shows two typical distribution of RTTs measured by a stationary UE.   It can be seen that the RTT distribution is multi-modal, and on further inspection the intervals between the first peak and the second is 9.6ms, the first and the third is 19.4ms, the first and the fourth is 28.5ms.   That is, the intervals between the peaks are 9.6ms, 9.8ms, 9.1ms.  These spacings are consistent across measurements at different times and locations and so seem to be a genuine feature of the LTE MAC.   Since the HARQ retransmit time in LTE is at least 8ms, one hypothesis is that the peaks correspond to packets transmitted with no retry, with two retries, with three retries etc.  It can be seen that on the link with lower signal quality (Figure \ref{fig:Link-layer-RTT-fit} (a)) there are additional peaks at around 73ms and 82ms and it is these which lead to the increase in the mean and variance of the RTT seen in Figure \ref{fig:RTT-mean-distribution}.

\begin{figure}
\centering
\begin{minipage}[b]{0.45\linewidth}
\scriptsize
\centering
\begin{tabular}{|c|cc|}
\hline
$i$ &  $\mu_i$ & $\sigma_i$ \\
\hline
1 &  $\mu_1$ & 0.02 \\
2 &  $\mu_1+9.6$ & 0.03 \\
3 &  $\mu_1+19.4$ & 0.04 \\
4 &  $\mu_1+28.5$ & 0.04\\
\hline
& \multicolumn{2}{|c|}{$w_i$}\\
\hline
1 & \multicolumn{2}{|c|}{$0.0079 CSQ -0.1285$}\\
2 &\multicolumn{2}{|c|}{$0.0430 CSQ -0.3834$}\\
3 &\multicolumn{2}{|c|}{$-0.0059 CSQ +0.3244$}\\
4 &\multicolumn{2}{|c|}{$-0.0096 CSQ +0.2937$}\\
\hline
\end{tabular}
\normalsize
(a) $\mu_i$, $\sigma_i$ and $w_i$
\label{fig:minipage1}
\end{minipage}
\quad
\begin{minipage}[b]{0.45\linewidth}
\centering
\includegraphics[width=\textwidth]{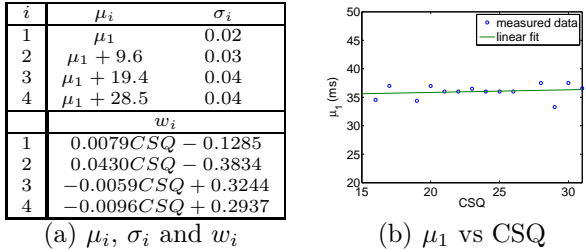}
(b) $\mu_1$ vs CSQ
\label{fig:minipage2}
\end{minipage}
\caption{Parameters of mixture of Gaussians fit to measured RTT distribution. }\label{fig:mix}
\end{figure}

Each peak in the RTT distribution can be well fitted by a mixture of Gaussian distributions $\sum_{i=1}^4 w_iN(\mu_i,\sigma_i)$, where $N(\mu,\sigma$ denotes a Gaussian distribution with mean $\mu$ and standard deviation $\sigma$, as indicated on Figure \ref{fig:Link-layer-RTT-fit}.   The mean and standard deviation parameters of the Gaussians in the mixture are quite consistent across runs in our data and are detailed in Figure \ref{fig:mix}.  The RTT $\mu_1$ at which the first peak occurs does vary across data sets, ranging from 32.5ms to 37.5ms, but this variation does not seem to be correlated with link quality, see Figure \ref{fig:mix}(b).   The main changes with CSQ are in the weights $w_i$, $i=1,\cdots, 4$ of the components of the mixture and linear fits for these are detailed in Figure \ref{fig:mix}(a).  This corresponds to a one-way latency of around 15ms, which seems to consistent with the LTE literature and is likely associated with a mix of wireless MAC delays and delay on the cellular wired backhaul

\begin{figure}
\centering
\subfloat[Stationary UE]{
\includegraphics[width=0.45\columnwidth]{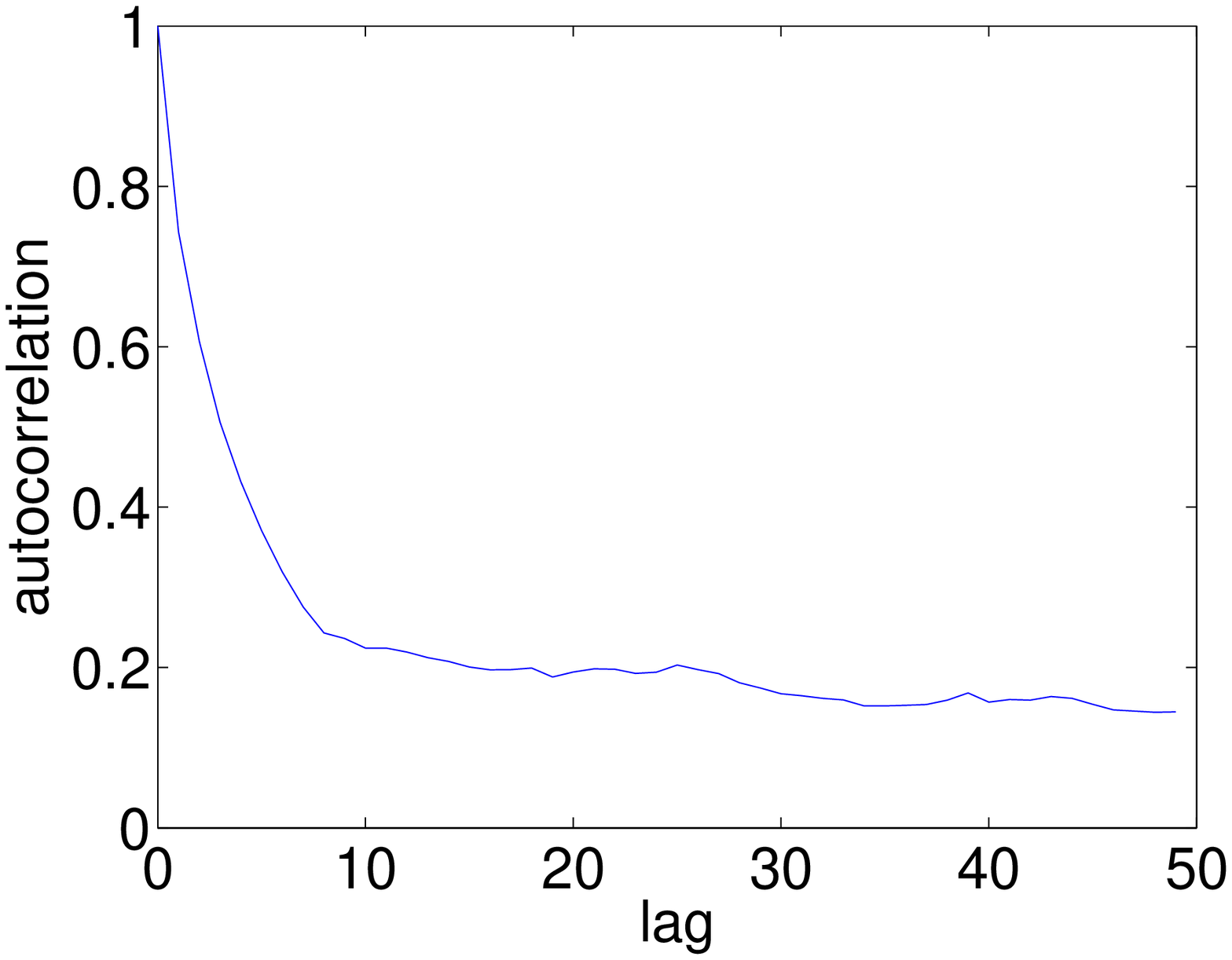}
}\quad
\subfloat[UE on moving tram.]{
\includegraphics[width=0.45\columnwidth]{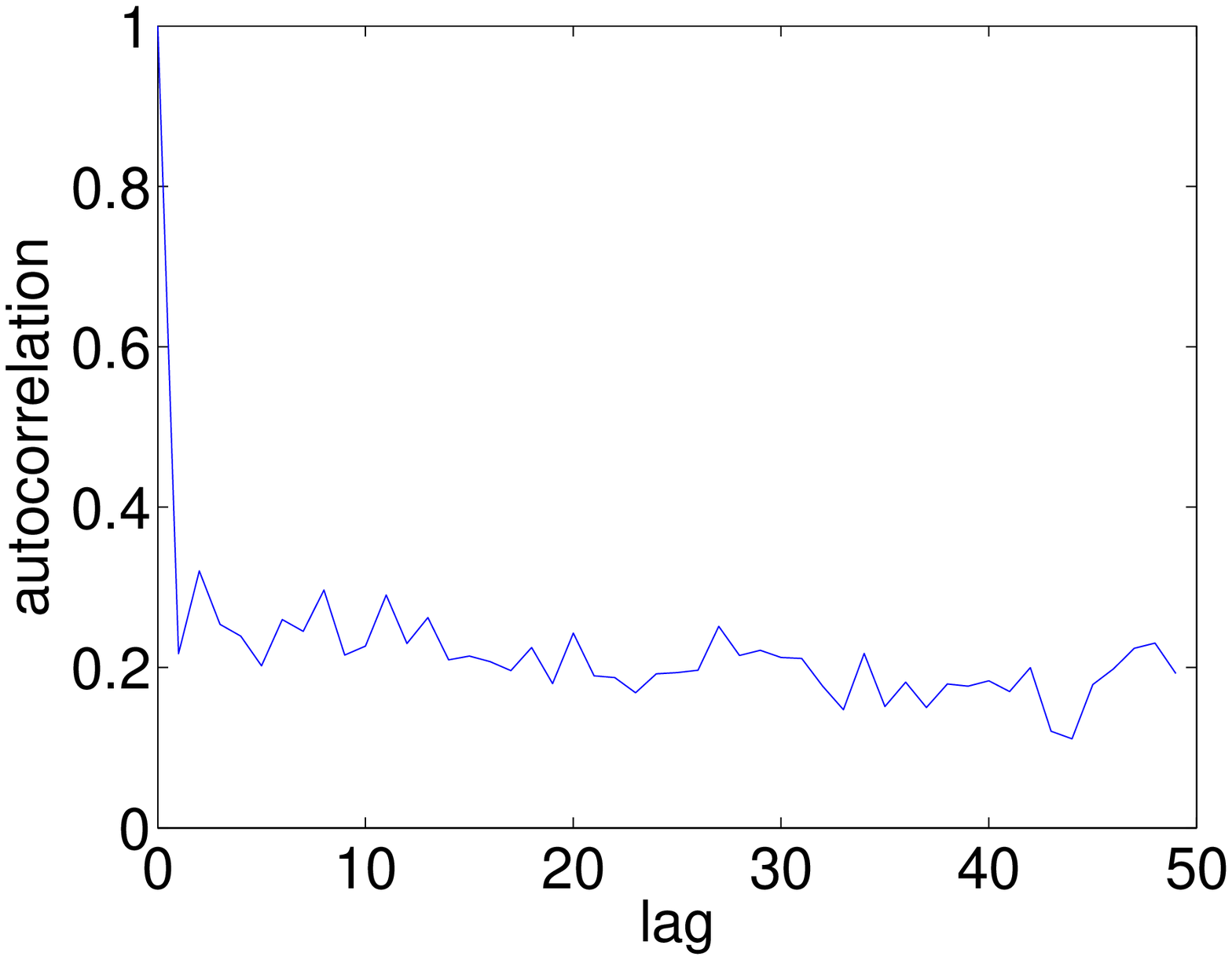}
}
\caption{Autocorrelation of measured RTT for a stationary UE and a moving UE.  Operator T, CSQ 19.  Stationary data corresponds to that in Figure \ref{fig:Link-layer-RTT-fit}(a). }\label{fig:RTTauto}
\end{figure}

Figure \ref{fig:RTTauto}(a)  plots the RTT autocorrelation.  It can be seen that the RTT values show evidence of correlation over intervals of about 10 samples.  However, the correlation decays quite quickly and so it is probably reasonable to model the RTTs as being i.i.d, at least to a first approximation. 

\subsection{RTT vs Link Quality: Operator M}

\begin{figure}
\centering
\subfloat[Distribution]{
\includegraphics[width=0.45\columnwidth]{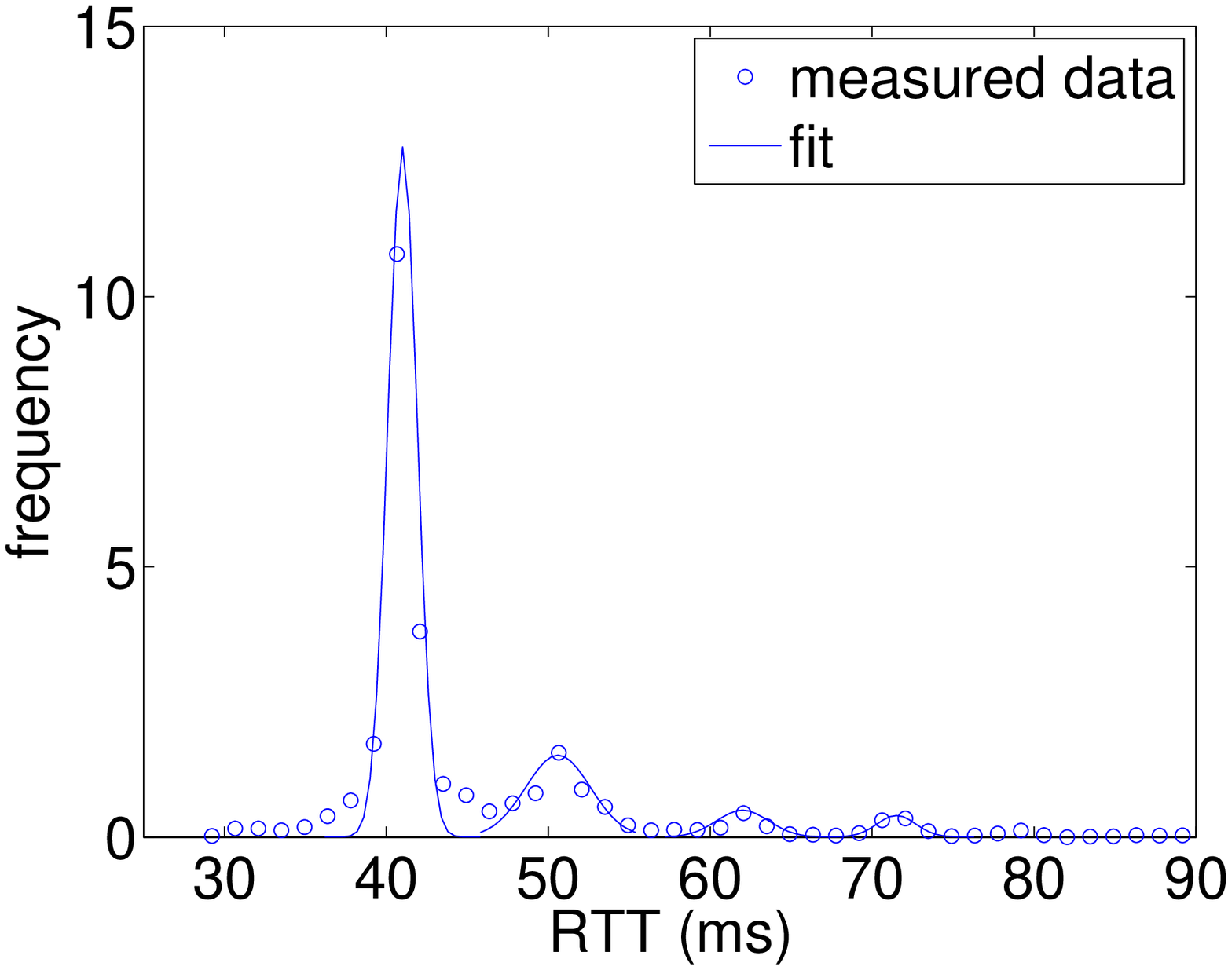}
}\quad
\subfloat[Autocorrelation]{
\includegraphics[width=0.45\columnwidth]{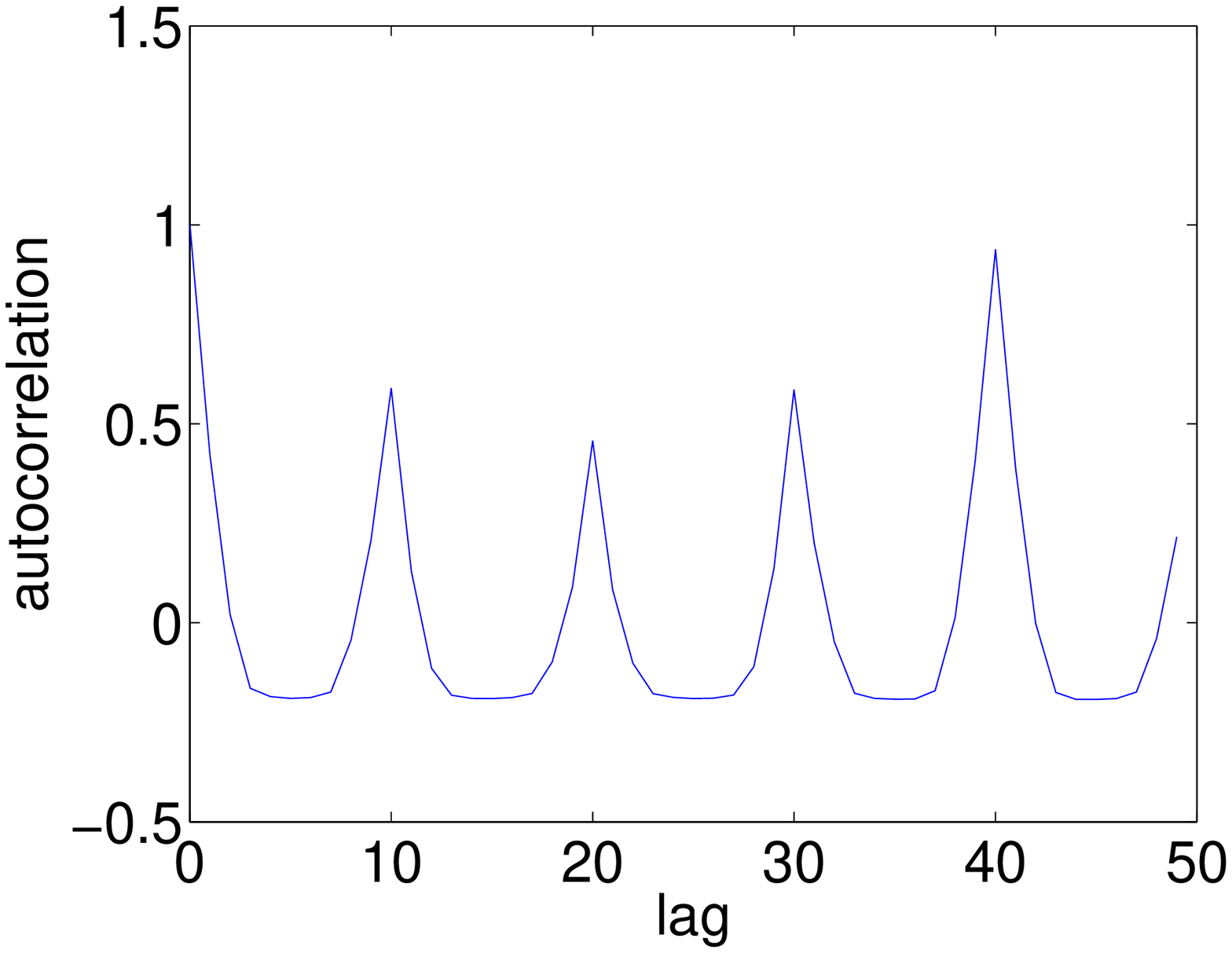}
}
\caption{\label{fig:Link-layer-RTT-fitM}Measured RTT for a stationary UE with CSQ of 27, operator M.  }
\end{figure}

We found that the measured RTT characteristics varied somewhat with the operator.  The foregoing data is for operator T.  When using operator M, 
Figure \ref{fig:Link-layer-RTT-fitM}(a) plots a typical measured RTT distribution.   It can be seen that, similarly to operator T, this distribution displays a number of peaks and can be approximated by a mixture of Gaussians (also indicated in the figure).   The intervals between the peaks are also similar to operator T, being 9.6ms, 11.4ms, 9.5ms.  However, the distribution is much more strongly concentrated around the first peak, and this is a consistent feature of measurements with operator M. 

Figure \ref{fig:Link-layer-RTT-fitM}(b) plots the autocorrelation of the data shown in Figure \ref{fig:Link-layer-RTT-fitM}(a).   It can be seen that the RTT measurements are essentially periodic, with period 10 samples.   This is in contrast to operator T where the RTT measurements show much less correlation.   We are unsure as to the source of this periodicity, but it is consistent across all measurements with operator M and so presumably reflects the scheduling strategy used.


\subsection{Impact of Mobility}
We also collected data for both operators on mobile UEs travelling on two tram lines within the city (data for both operators was collected simultaneously using two UEs).   For both operators this data showed very similar RTT mean and standard deviation vs link quality behaviour to the above data for a stationary UE.   RTT data for both operator exhibits low autocorrelation, see Figure \ref{fig:RTTauto}(b) for the autocorrelation with operator T.   Note that for operator M this is unlike the situation for a stationary UE, where the RTTs are observed to be essentially periodic.

\begin{figure}
\centering
\subfloat[CSQ 19]{
\includegraphics[width=0.45\columnwidth]{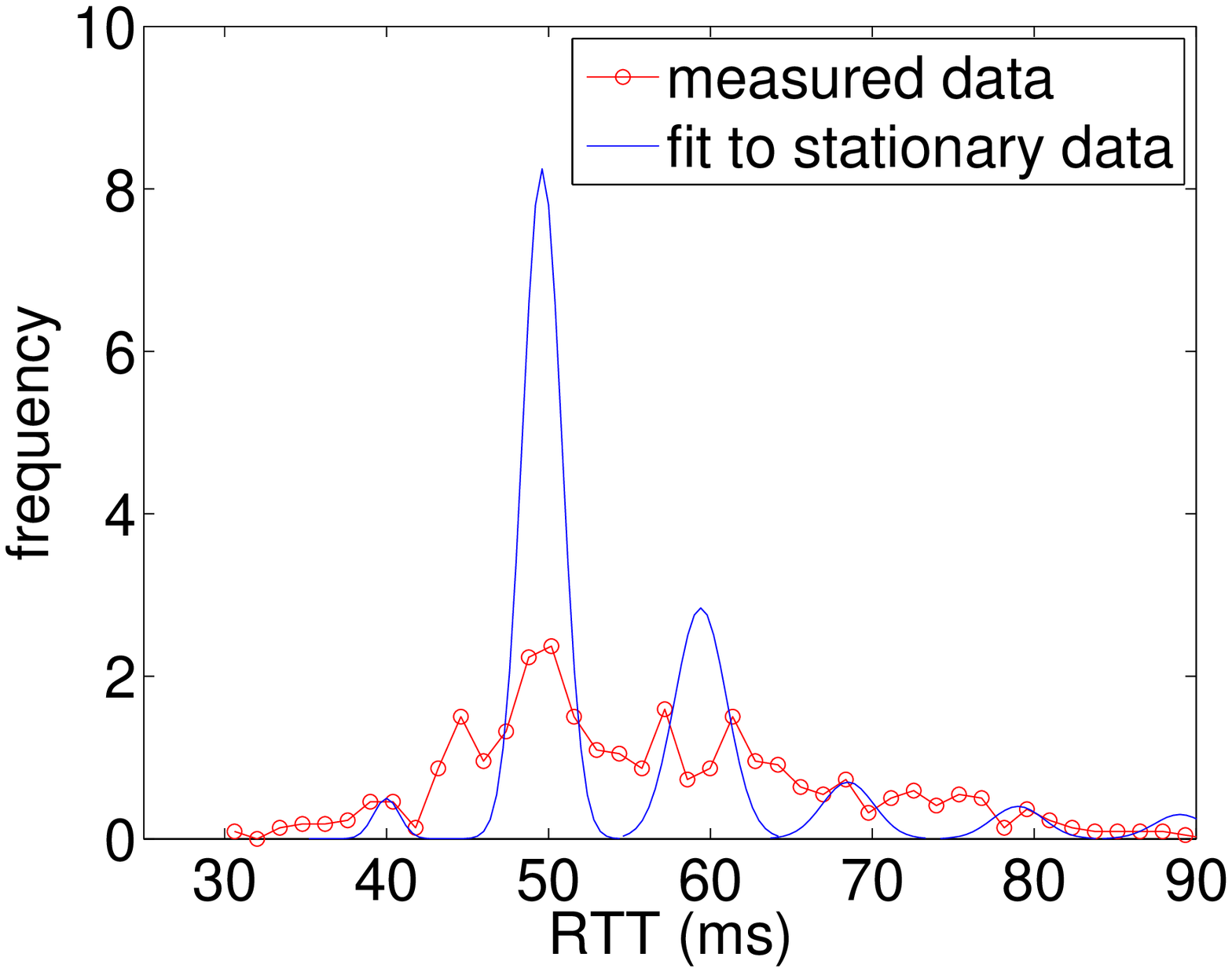}
}\quad
\subfloat[CSQ 31]{
\includegraphics[width=0.45\columnwidth]{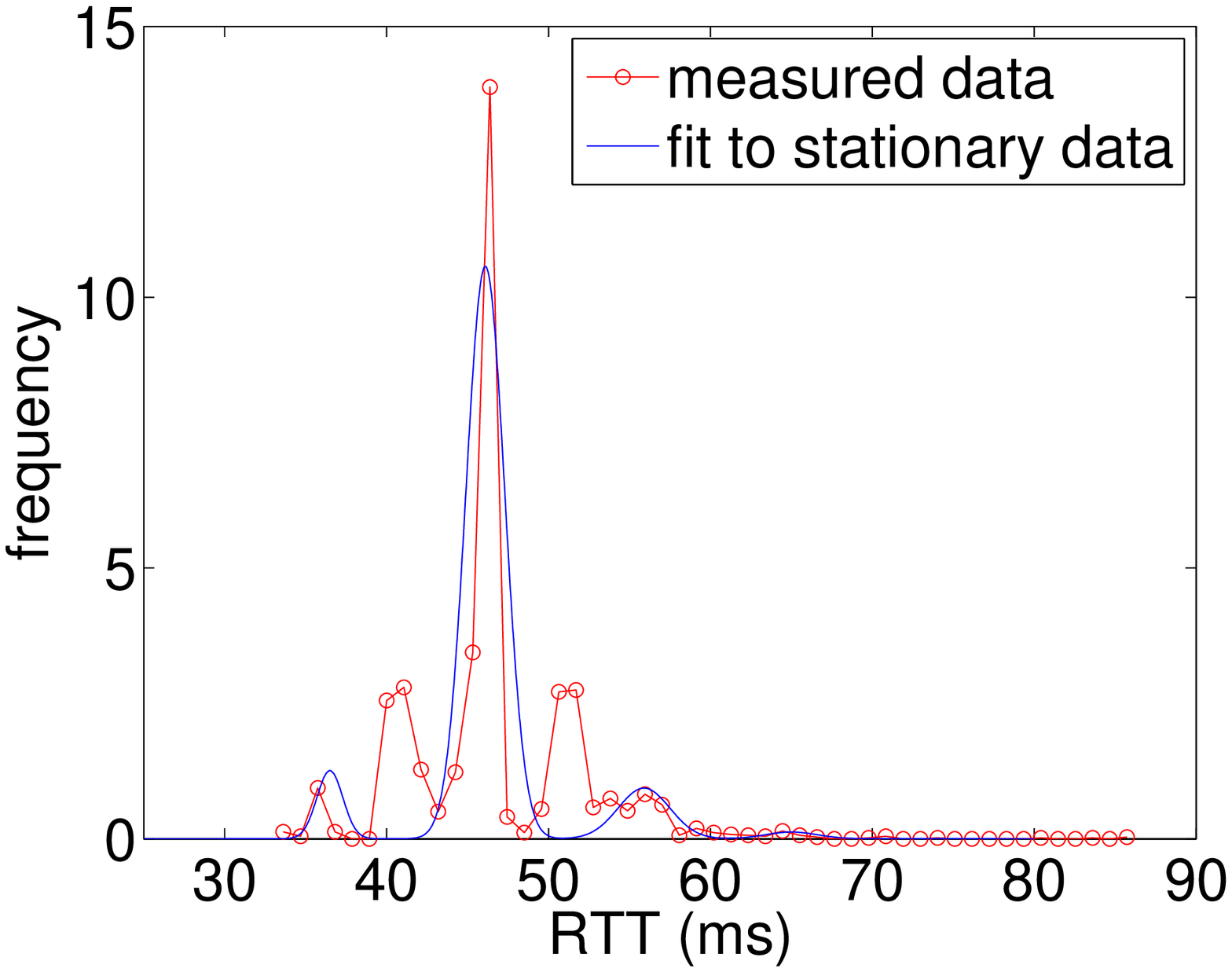}
}
\caption{\label{fig:Link-layer-RTT-fitTram}Measured RTT for a moving UE, operator T.  }
\end{figure}

In addition to removing the periodicity from operator M RTTs, mobility also changes the RTT distribution.   Figure \ref{fig:Link-layer-RTT-fitTram} plots the measured RTT distribution for a moving UE with operator T and links CSQs of 19 and 31.    Also shown is the fit to the RTT distribution for a stationary UE with operator T and link CSQs of 19 and 31, taken from Figure \ref{fig:Link-layer-RTT-fit}.   For a CSQ of 31 (a high quality link), the RTT distribution is much the same as in the stationary measurements.   However, for a CSQ of 19, the RTT distribution is somewhat ``smeared out'' for the moving UE.   For the moving UE the plot here is generated by taking all RTT measurements for which the CSQ is 19.   While RTT measurements are taken at 50ms intervals, the CSQ is only updated once per second, and since the CSQ is changing as the tram moves some of the RTT data plotted may in fact belong to a nearby CSQ value.  As CSQs increase above 19 the RTT distribution changes the weights on the different peaks, see Figure \ref{fig:Link-layer-RTT-fit}, and  so this mixing of data may explain Figure \ref{fig:Link-layer-RTT-fitTram}(a).

\section{Bandwidth} \label{sec:bandwidth}

\subsection{Bandwidth vs Link Quality: Operator T}

\begin{figure}
\centering
\subfloat[Mean]{
\includegraphics[width=0.45\columnwidth]{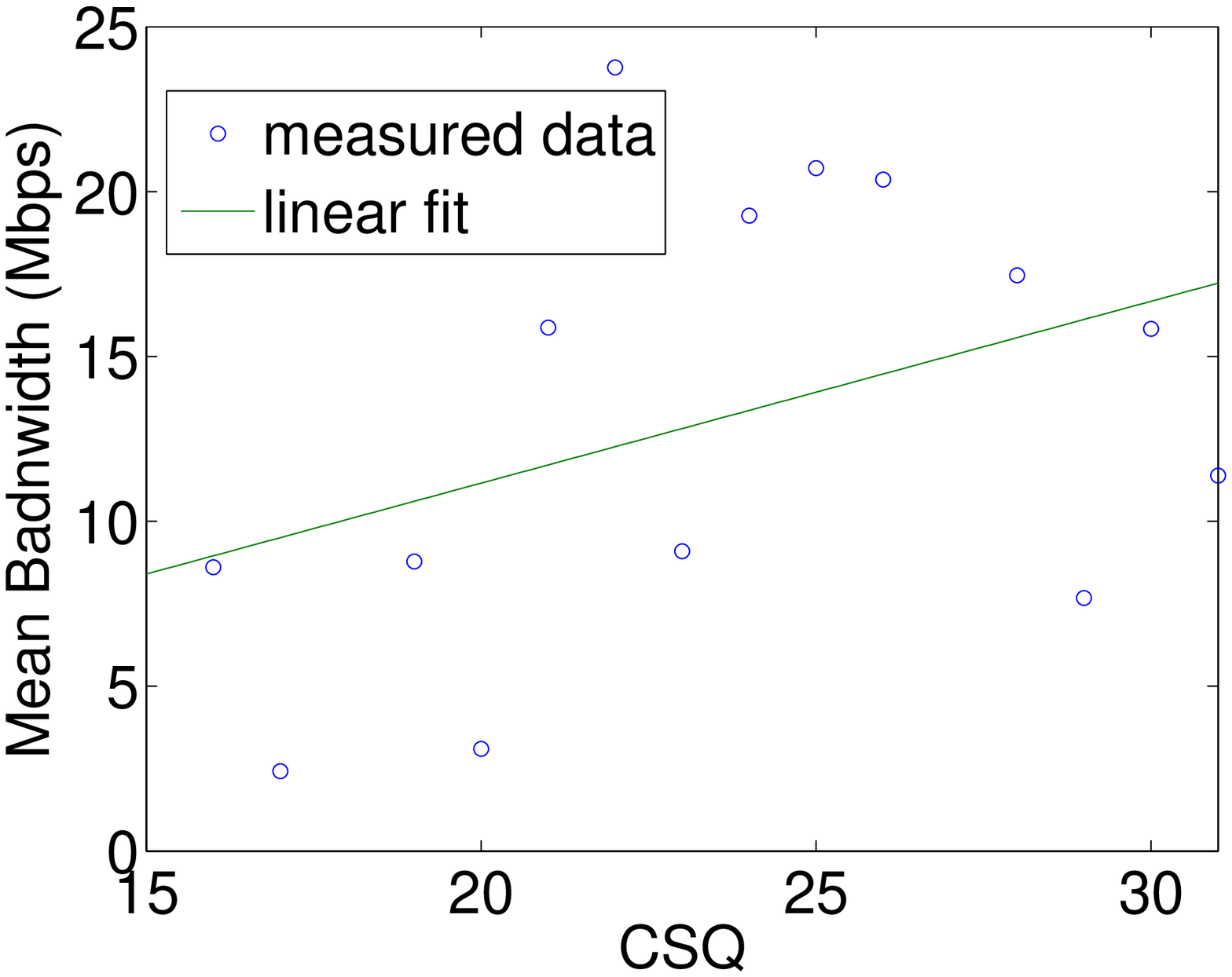}
}\quad
\subfloat[Standard deviation]{
\includegraphics[width=0.45\columnwidth]{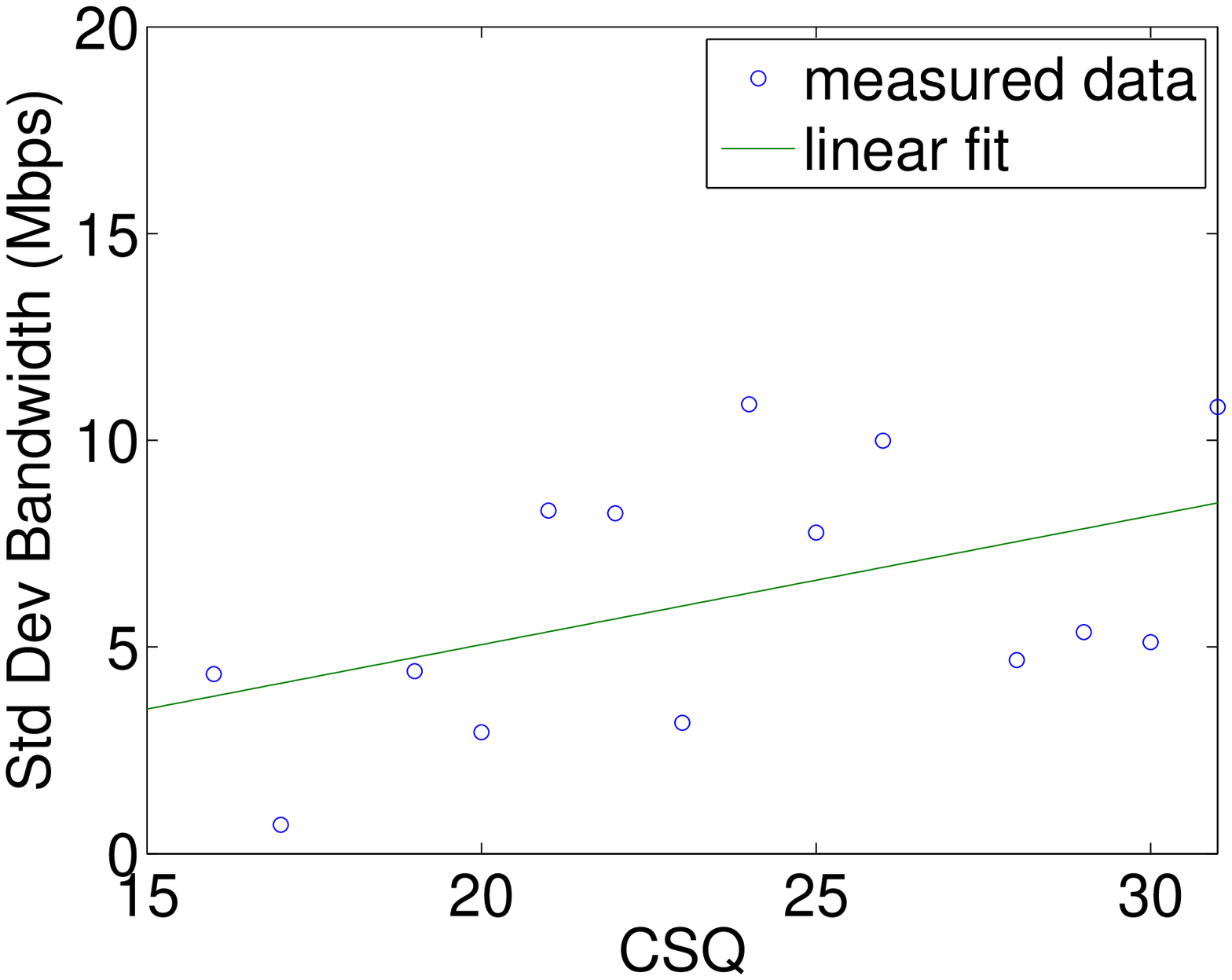}
}
\caption{\label{fig:Bandwidth-mean-distribution}Mean and standard deviation of measured bandwidth vs CSQ, plus linear fits. Operator T.}
\end{figure}

Figure \ref{fig:Bandwidth-mean-distribution} plots the mean and standard deviation of the bandwidth measured by a stationary UE at locations with a range of link qualities.   It can be seen that the mean bandwidth increases with link quality, rising from around 10Mbps for links with a CSQ of 15 to around 20Mbps for links with a CSQ $> 30$.   

\begin{figure}
\centering
\subfloat[Distribution]{
\includegraphics[width=0.45\columnwidth]{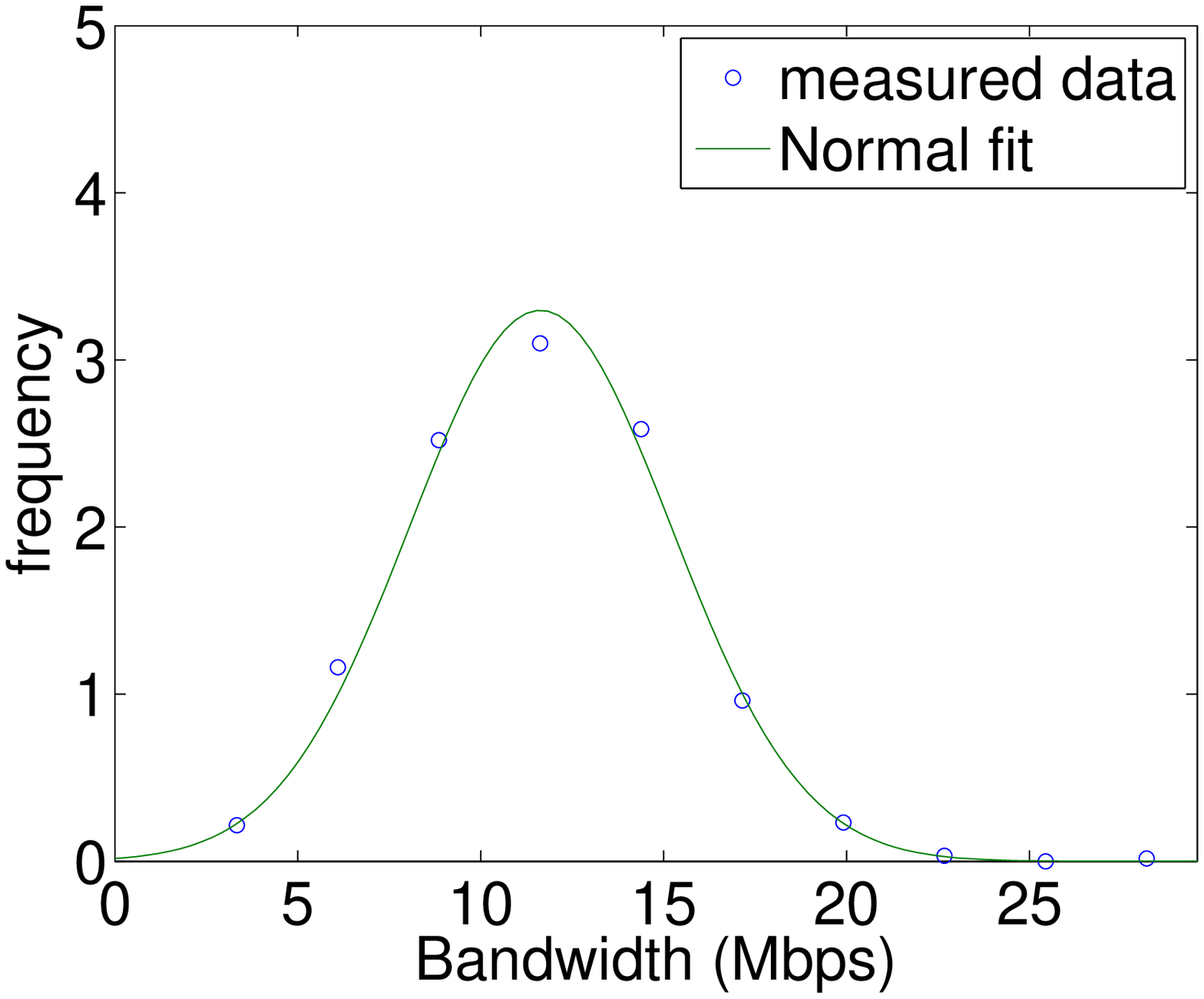}
}
\subfloat[Autocorrelation]{
\includegraphics[width=0.45\columnwidth]{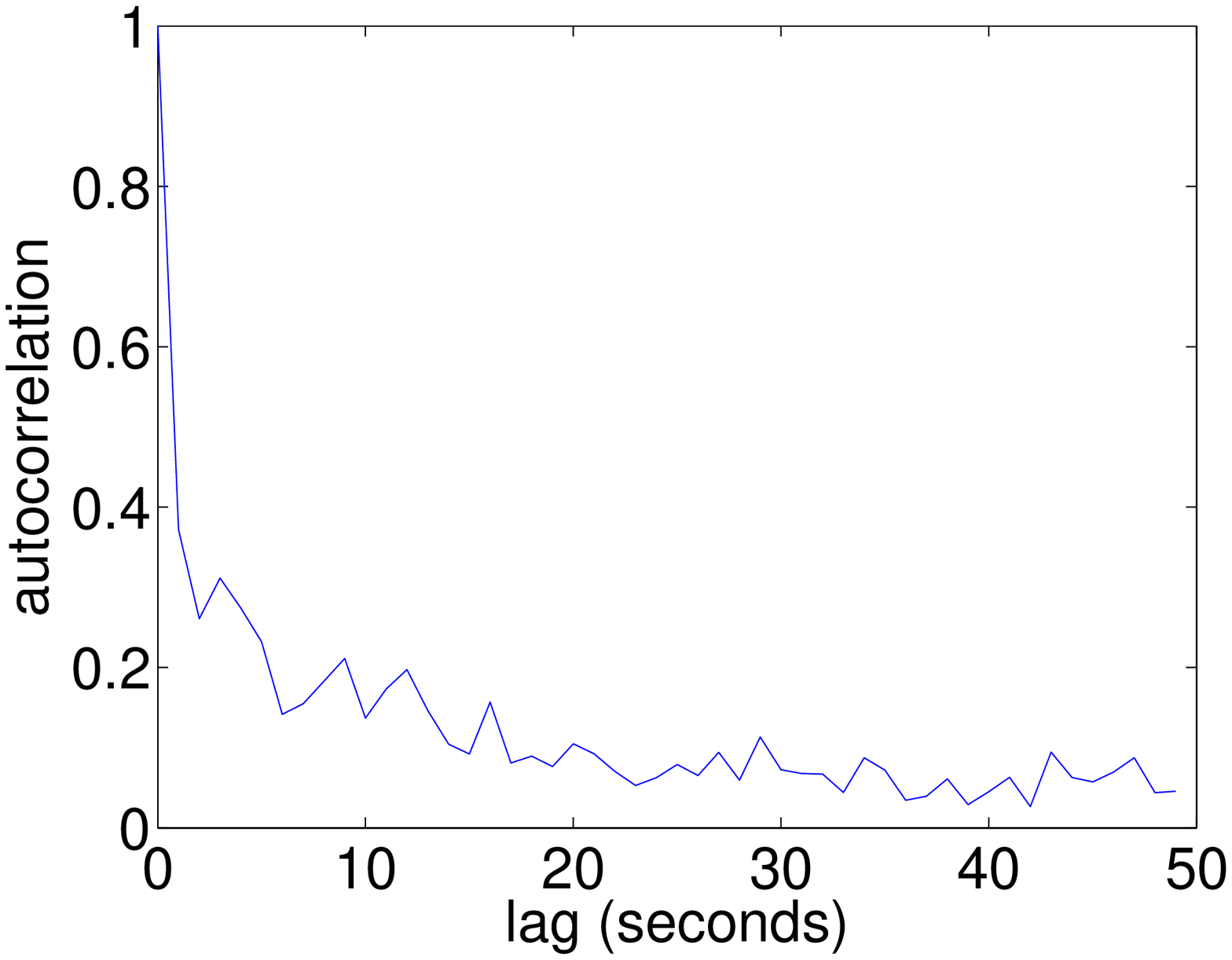}
}
\caption{\label{fig:Bandwidth-vs-CSQ}Measured bandwidth.  Data shown in both plots is for the same link with CSQ 31.  Operator T.  }
\end{figure}

In more detail, Figure \ref{fig:Bandwidth-vs-CSQ}(a) plots the distribution of bandwidth measured on a link with CSQ=31, and also a Gaussian distribution fitted to this data.   It can be seen that a  Gaussian distribution, truncated to be non-negative, provides a reasonable fit to the bandwidth measurements.  It can be seen from Figure \ref{fig:Bandwidth-vs-CSQ} that the mean and standard deviation variation with CSQ can be approximately modelled as $\mu=0.55\times CSQ+0.13$ for the mean and $\sigma=0.31\times CSQ-1.17$ for the standard deviation.    Figure \ref{fig:Bandwidth-vs-CSQ}(b) plots the autocorrelation of the measured bandwidth.  It can be seen that although the bandwidth values show more correlation than the RTT data, the correlation is still quite weak and so the bandwidth probably reasonably modelled as i.i.d.

\begin{figure}
\subfloat[CSQ=30]{\includegraphics[width=0.45\columnwidth]{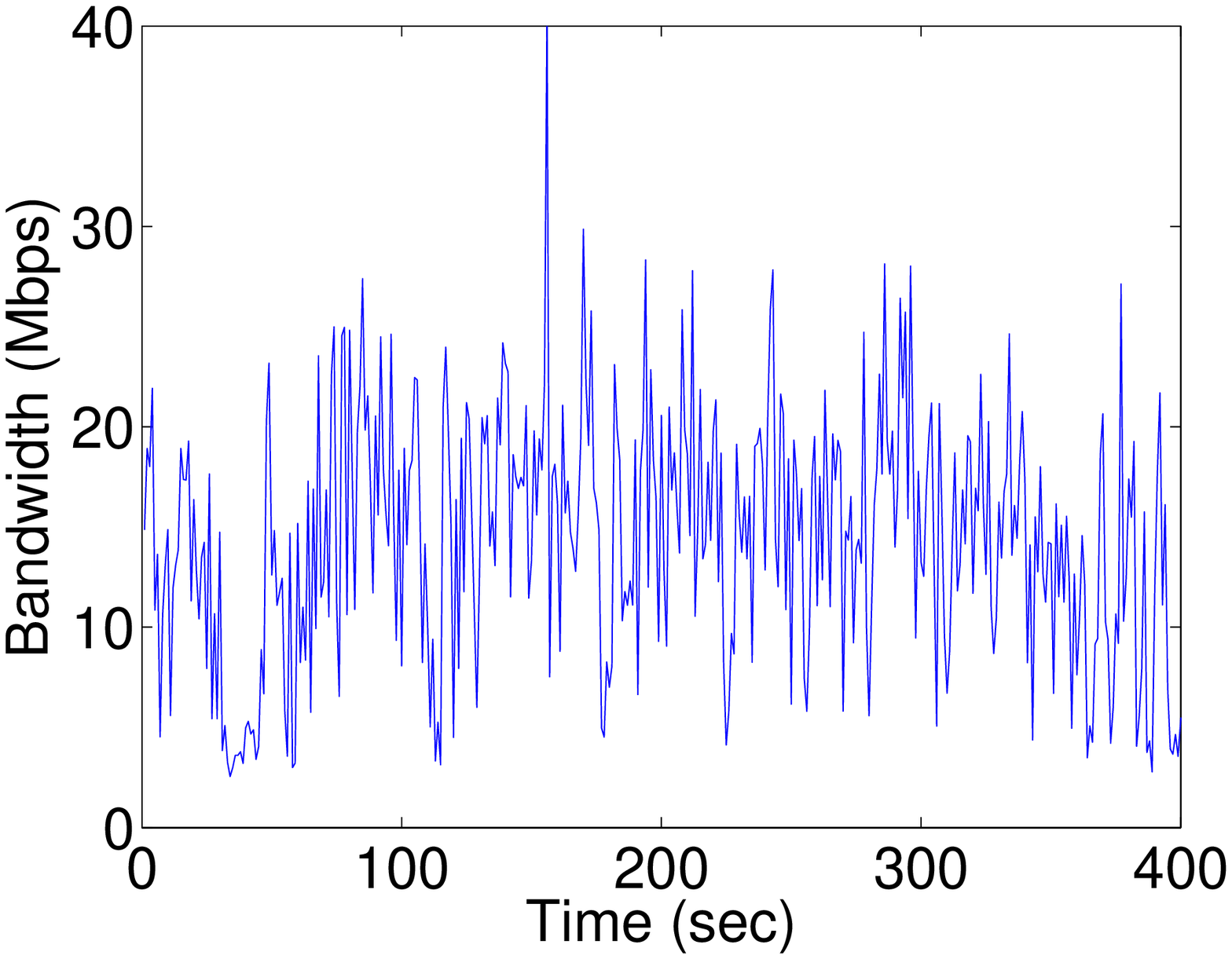}
}\quad
\subfloat[CSQ=18]{\includegraphics[width=0.45\columnwidth]{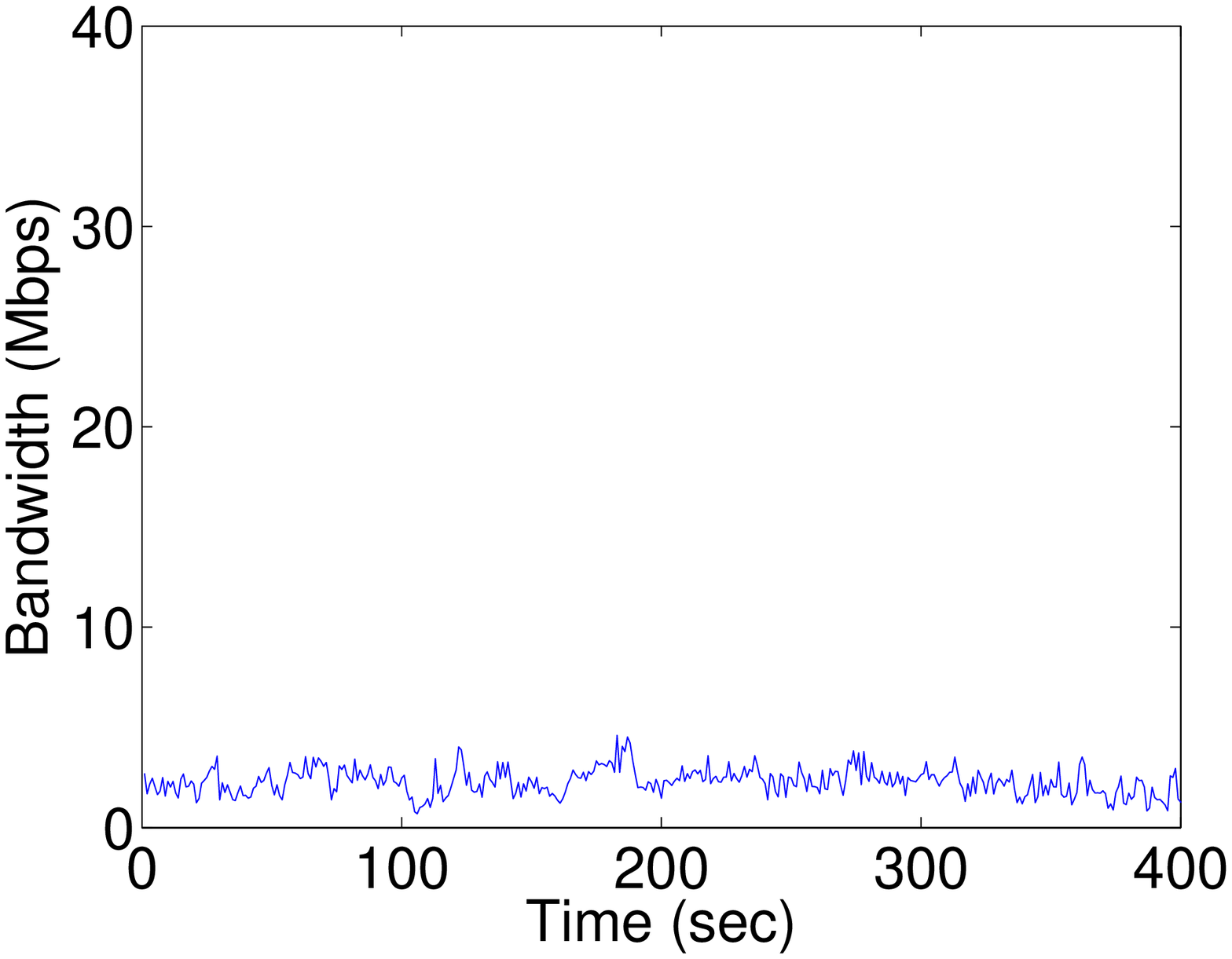}
}
\caption{\label{fig:Variability-of-bandwidth}Time histories of measured bandwidth.  Stationary UE, operator T.}
\end{figure}

Observe from Figure \ref{fig:Bandwidth-mean-distribution}(b) that the standard deviation of the bandwidth increases as the link quality increases, which is quite surprising (we might have expected variability to decrease with increasing link quality if it is associated with link layer retransmissions to recover from loss).   This effect is also evident in Figure \ref{fig:Variability-of-bandwidth} which shows two example time histories of measured bandwidth, one at a location where the link quality is good (CSQ is 30) and one where the link quality is poorer (CSQ 18).     We note that, although it is hard to confirm the level of cell load, we believe that these measurements are for lightly loaded cells and so the variability is \emph{not} due to changing cell traffic load.  The source of this variability is not clear at present, but we suspect that it is associated with the time granularity of scheduling updates by the cellular basestation since we note that the ratio of the standard deviation to the mean bandwidth is approximately constant with CSQ (assuming changes in bandwidth are mainly due to changes in the modulation and coding scheme used, then timing granularity would result in bandwidth fluctuations that scale with bandwidth, as observed).   

\subsection{Bandwidth vs Link Quality: Operator M}
\begin{figure}
\centering
\subfloat[Mean]{
\includegraphics[width=0.45\columnwidth]{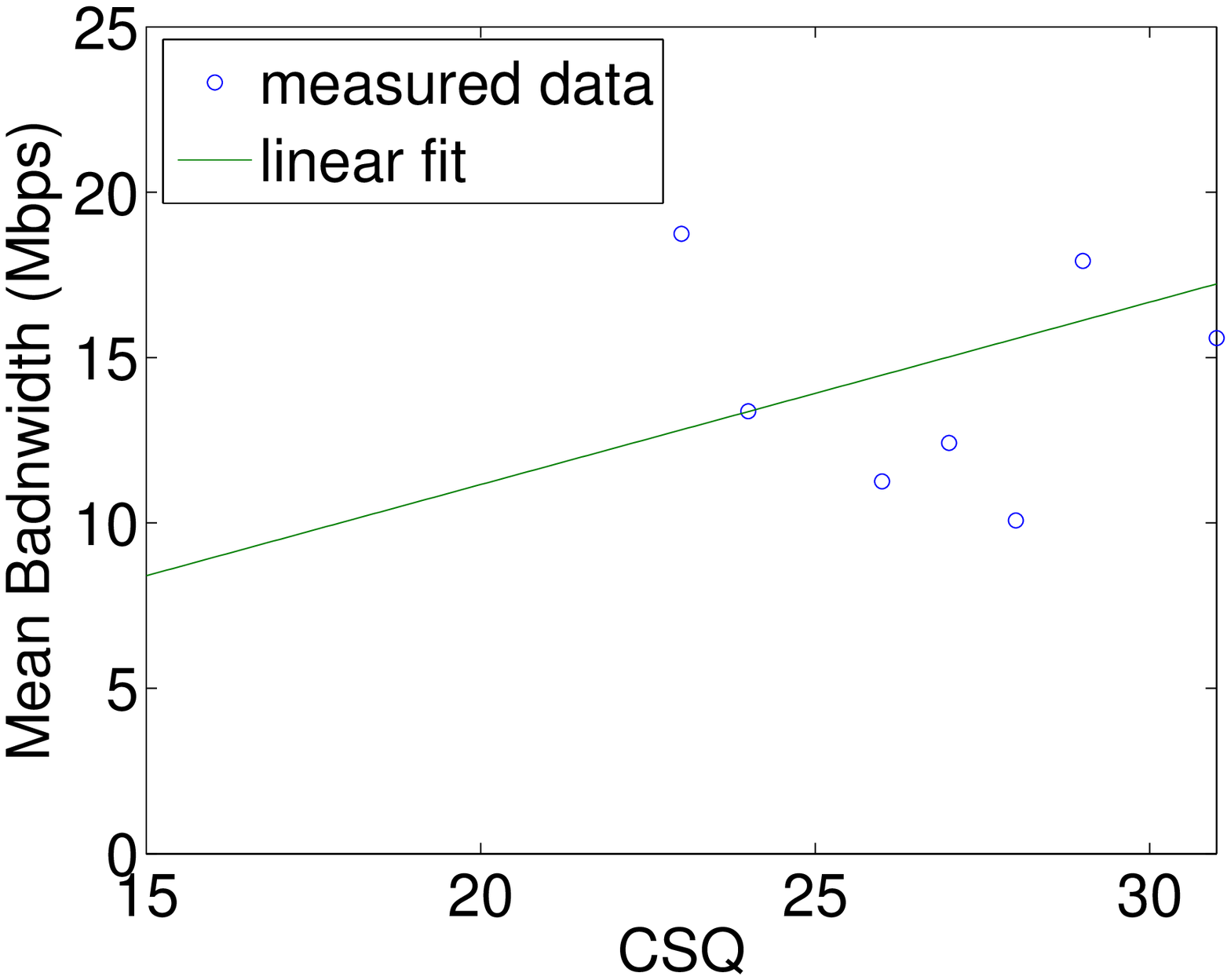}
}\quad
\subfloat[Standard deviation]{
\includegraphics[width=0.45\columnwidth]{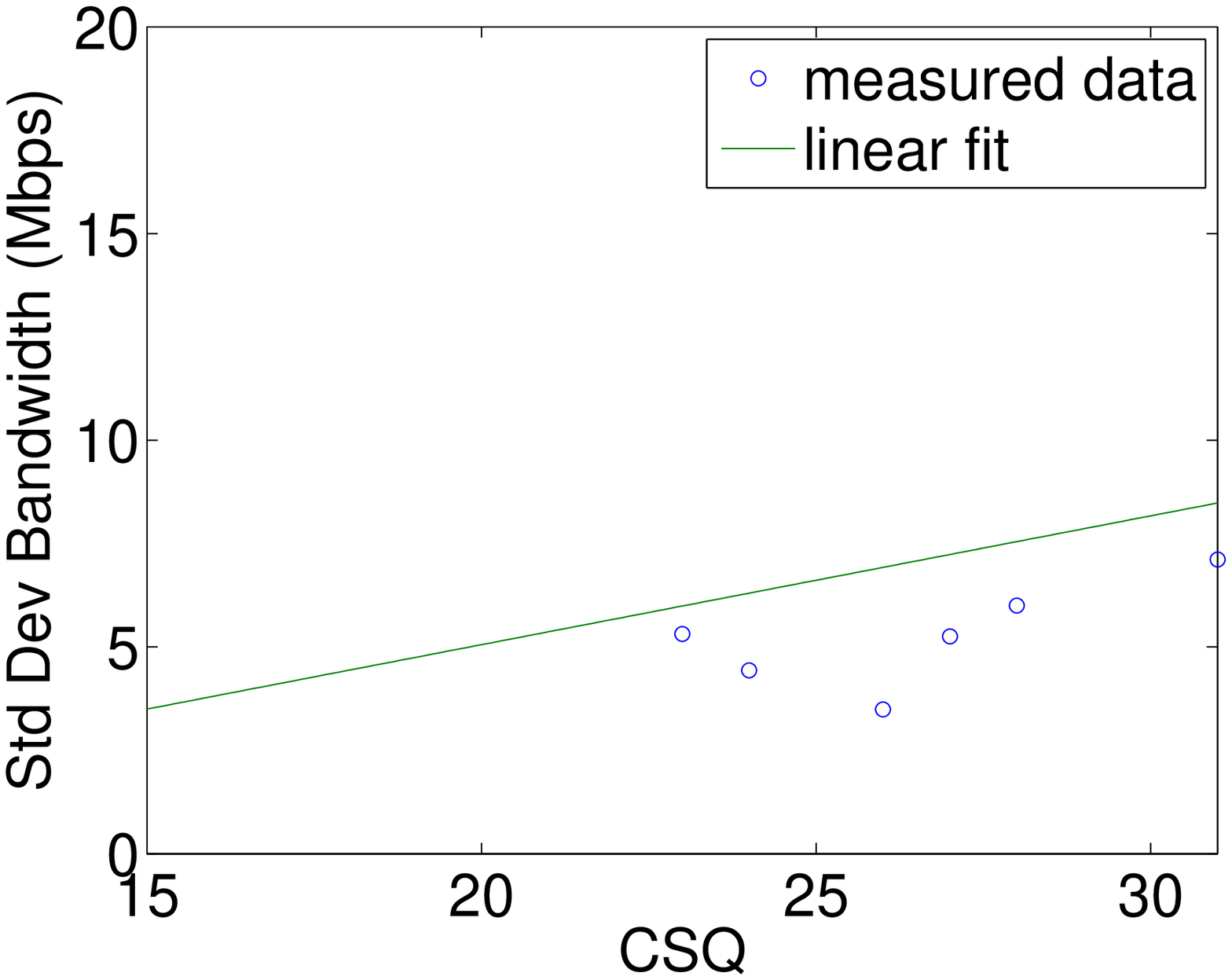}
}
\caption{\label{fig:Bandwidth-mean-distributionM}Mean and standard deviation of measured bandwidth vs CSQ, plus linear fits. Operator M.}
\end{figure}

\begin{figure}
\centering
\subfloat[Distribution, CSQ=31]{
\includegraphics[width=0.45\columnwidth]{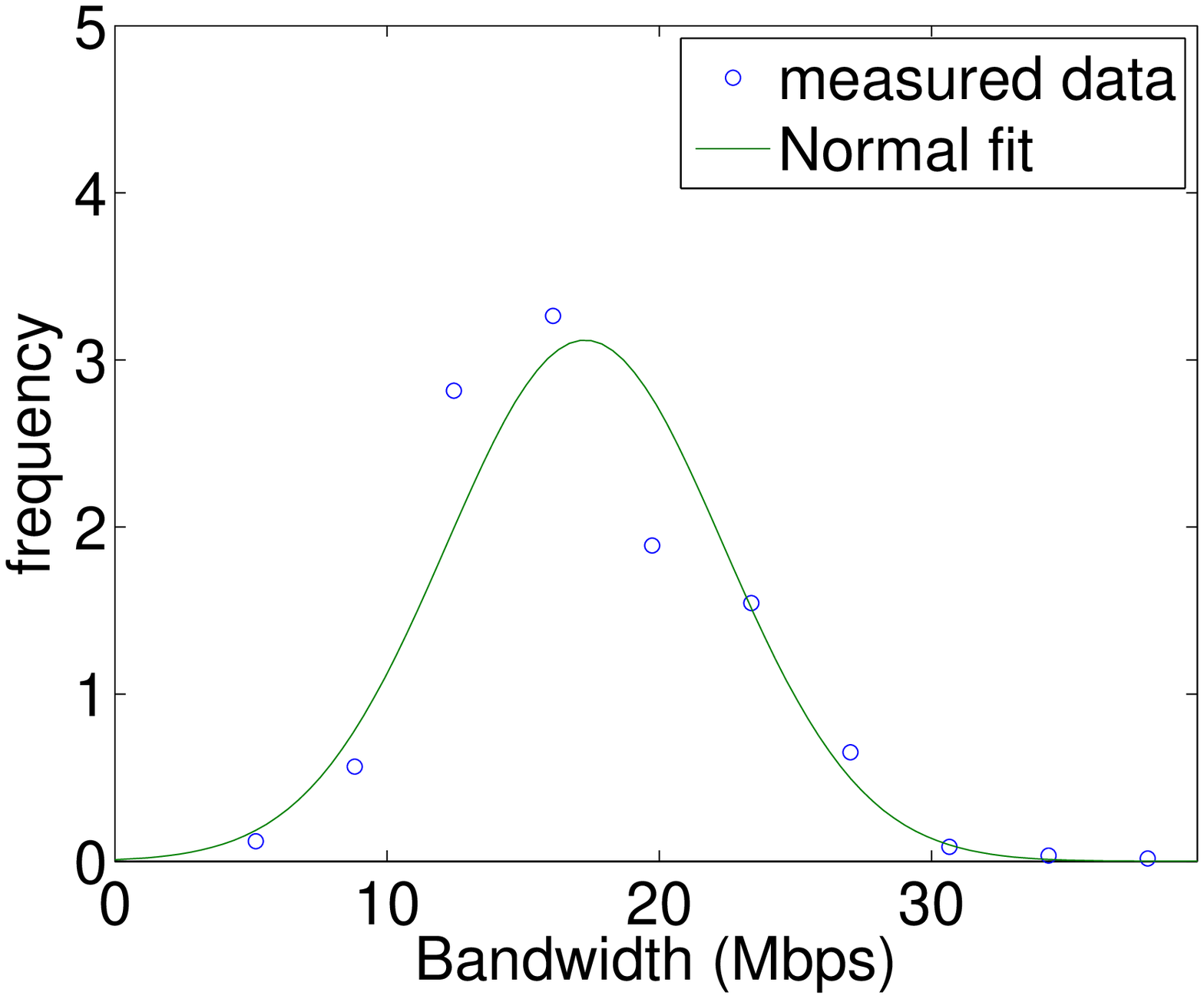}
}
\subfloat[Autocorrelation]{
\includegraphics[width=0.45\columnwidth]{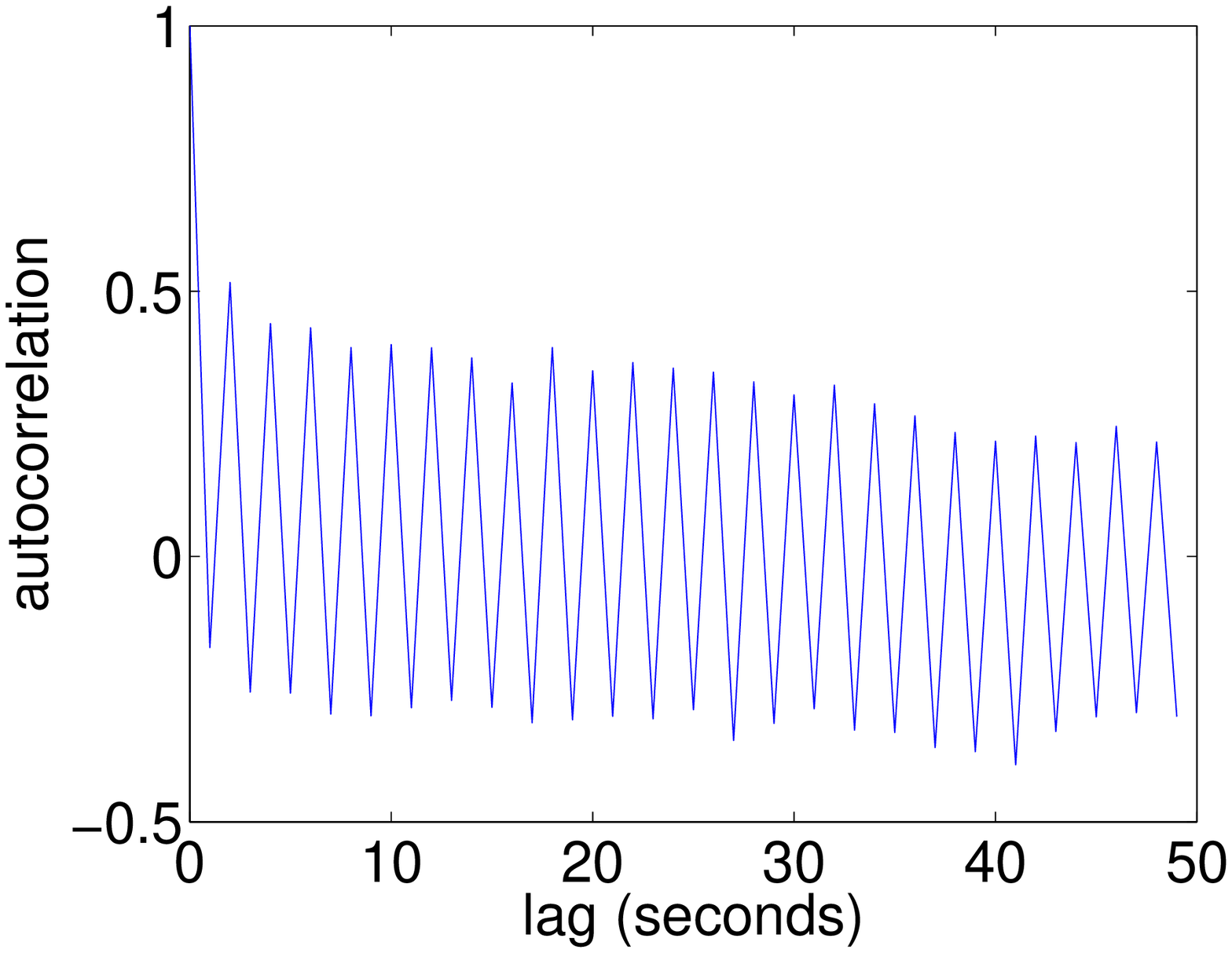}
}
\caption{\label{fig:Bandwidth-vs-CSQM}Measured bandwidth.  Data shown in both plots is for the same link with CSQ 31.  Operator M.  }
\end{figure}

Figure \ref{fig:Bandwidth-mean-distributionM} plots the mean and standard deviation of the measured bandwidth vs CSQ for operator M.  Also shown on the plots are the linear fits from operator T (i.e. the lines are the same as on Figure \ref{fig:Bandwidth-mean-distribution}).   It can be seen that the behaviour is similar to that of operator T.  With operator M, the distribution of bandwidth values is also well approximated by a Gaussian distribution, see Figure \ref{fig:Bandwidth-vs-CSQM}(a).  However, perhaps unsurprisingly in light of the periodicity observed in the RTT data for this operator, the bandwidth measured shows significant periodicity, see Figure \ref{fig:Bandwidth-vs-CSQM}(b).   

%
%

\subsection{Impact of Mobility}
Similarly to the case with RTT,  for both operators bandwidth data collected using mobile UEs travelling on two tram lines within the city showed very similar behaviour to the above data for a stationary UE.   

\section{Empirical LTE Model}

It can be seen from this measured data that a relatively simple empirical model of LTE RTT and bandwidth can be obtained, as follows.   

\begin{enumerate}
\item For uplink and downlink there is per UE queueing, so queueing delay is decoupled between UEs sharing the same cell.    

\item RTT is modelled as an i.i.d mixture of Gaussians  $\sum_{i=1}^4 w_iN(\mu_i,\sigma_i)$, where for a given CSQ the parameters $w_i$, $\mu_i$, $\sigma_i$ are given by Figure \ref{fig:mix}.

\item Bandwidth is modelled as i.i.d Gaussian distributed, with distribution parameters $\mu=0.55\times CSQ+0.13$ and $\sigma=0.31\times CSQ-1.17$.   That is, to generate a realistic sequence of bandwidth values for a link with a given CSQ value, draw a sequence of i.i.d Gaussian values distributed with pdf $e^{-(x-\mu)^2/(2\sigma^2)}/(\sigma\sqrt{2\pi})$ where $\mu=0.55\times CSQ+0.13$ and $\sigma=0.31\times CSQ-1.17$.
\end{enumerate}

The LTE link layer is complex and contains numerous design parameters the choice of which is left proprietary.  The complexity makes simulation computationally demanding, while the many unspecified design parameters make selection of realistic configurations problematic.   Since the above empirical model is much simpler than detailed LTE simulation models, it should be more convenient for evaluation of application layer performance and for evaluation over longer time scales (where simulation would take too long).   Since the model is based on measurements from actual LTE network deployments, it evidently reflects realistic configurations.
 
\section{Summary} \label{sec:summary}
We present results from an extensive LTE measurement campaign in Dublin, Ireland using a custom performance measurement tool.   Performance data was measured at a variety of locations within the city (including cell edge locations, indoors, outdoors etc) as well as for mobile users on public transport within the city.   Using this data we derive a model of the characteristics of link layer RTT and bandwidth vs link signal strength.   This model is suited to use for performance evaluation of applications and services, and since it is based on real measurements it allows realistic evaluation of performance.

\begin{figure}
\centering
\subfloat[RTT]{
\includegraphics[width=0.45\columnwidth]{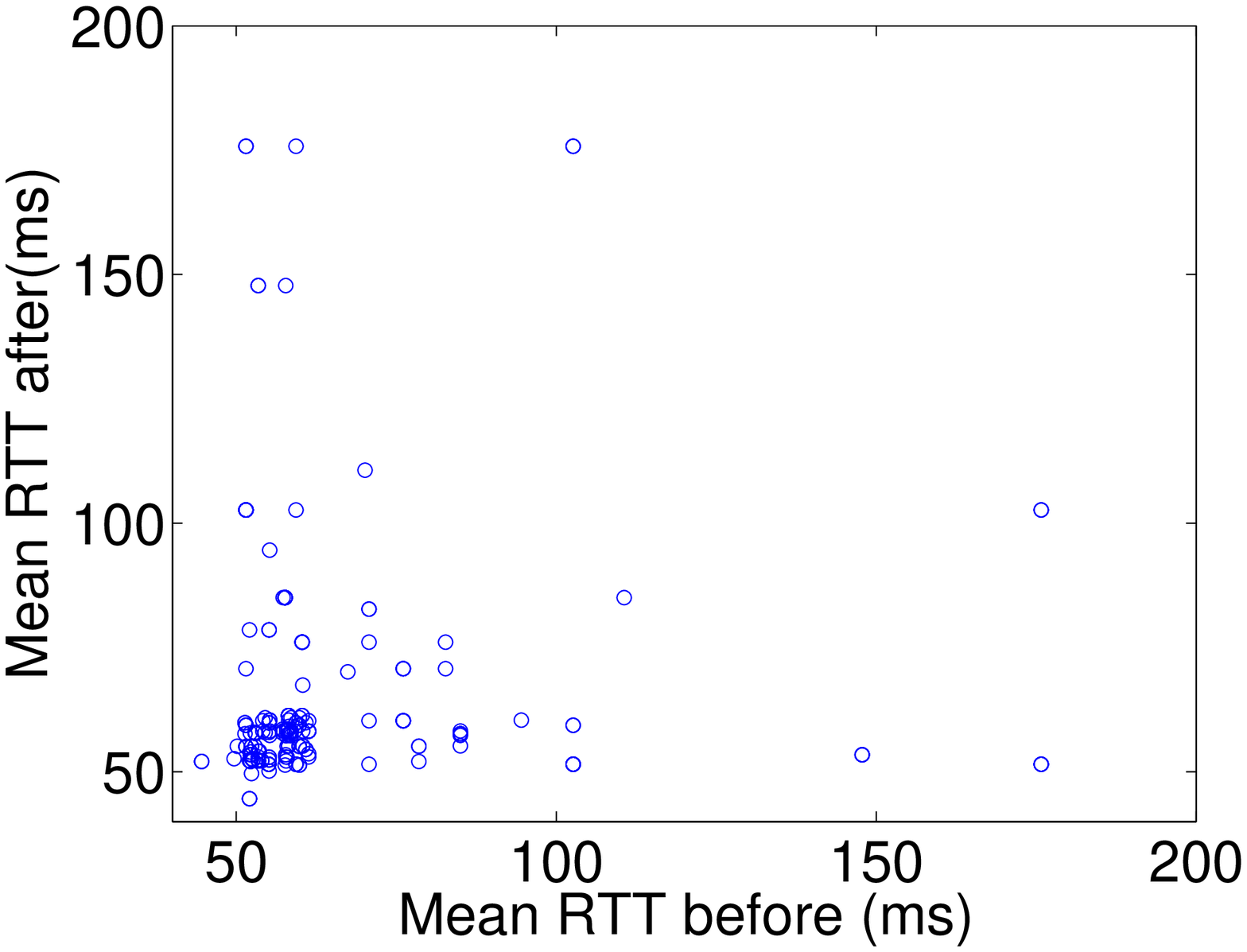}
}
\subfloat[CSQ]{
\includegraphics[width=0.45\columnwidth]{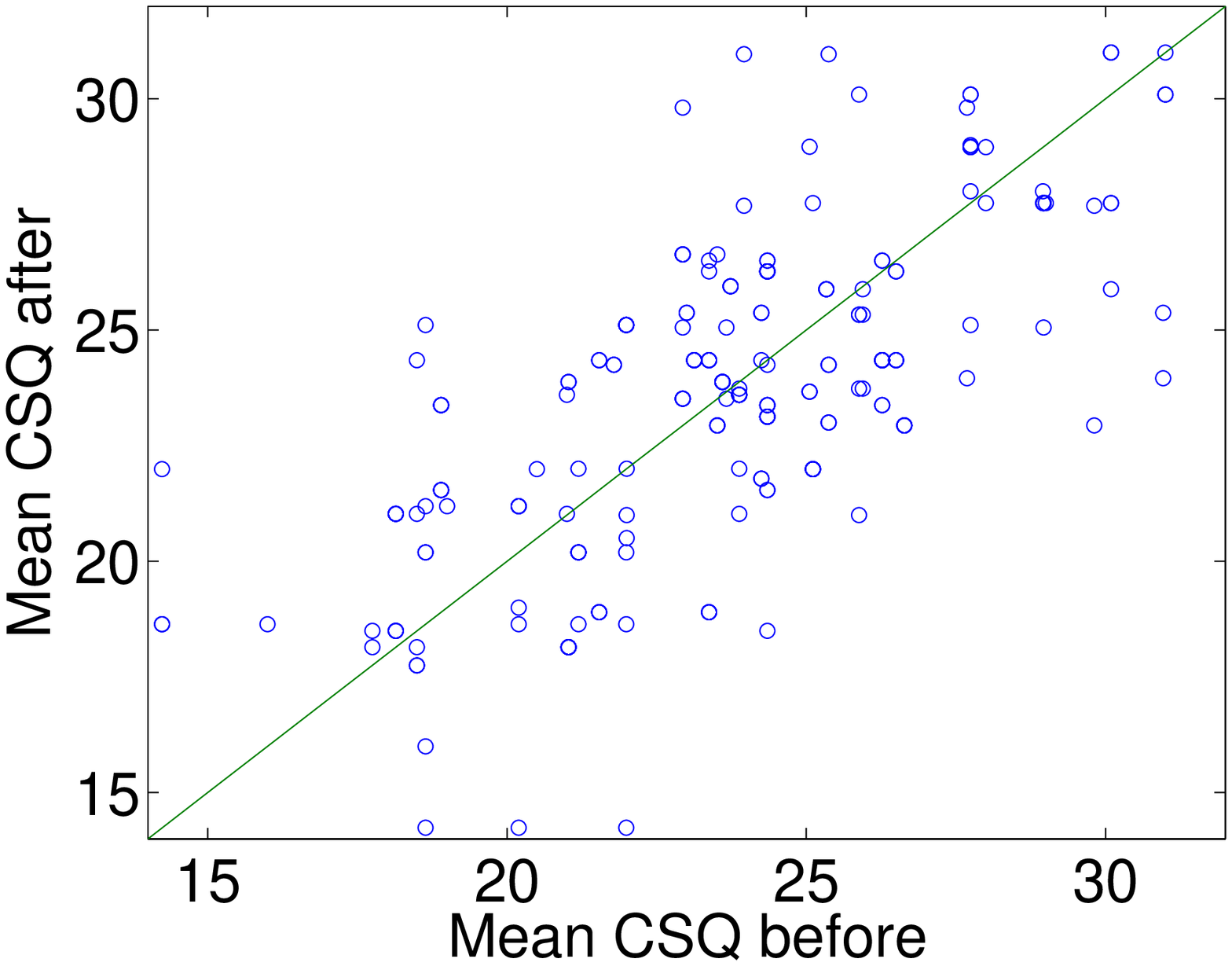}
}
\caption{\label{fig:handover}Measured RTT and CSQ before and after handover. UE travelling on tram, operator M. }
\end{figure}

While our measurement campaign provided extensive data on handover, we leave modelling of handover between cells to future work owing to lack of space.  We do, however, note briefly that the handover strategy used by an operator can have a significant impact on performance and can vary between operators.   For example, Figure \ref{fig:handover}(a) plots the mean RTT measured by a moving UE (travelling on a tram) before cell handover and afterwards.   Figure \ref{fig:handover}(b) plots the corresponding CSQ before and after handover.  It can be seen from Figure \ref{fig:handover}(b) that handovers tend to occur to cells with similar CSQ, as might be expected.  From Figure \ref{fig:handover}(a) we can see that the mean RTT before and after handover tends to be bunched around 50ms, in line with our other measurements above.  However, Figure \ref{fig:handover}(b) also shows that sometimes handover greatly increases the mean RTT, \emph{e.g.} from 50ms to 180ms.   Such a large mean RTT is indicative of a problematic link and we observe that handovers to more poorly performing cells need not be infrequent. \newline  

\bibliographystyle{plain}
\bibliography{therefs}

\end{document}